\documentclass{article}
\usepackage{iclr2027_arxiv_no_line,times}

\iclrfinalcopy

\usepackage{setspace,amsmath,amssymb,bm,theorem,graphicx,epstopdf,algorithm,float,color,mathtools,physics}
\usepackage[table,xcdraw]{xcolor}
\usepackage{subcaption}
\usepackage{booktabs}
\usepackage{tabularx}
\usepackage{algorithm,algorithmic}
\usepackage{dsfont}
\usepackage[short,c2]{optidef}
\usepackage{hyperref}
\usepackage{url}
\hypersetup{
  pdftitle={The Hidden Perception Constraint in Task-Aware Compression},
  pdfauthor={Sahan Liyanaarachchi, Semih Akkoc, Sennur Ulukus, Aylin Yener}
}
\usepackage{graphicx}
\usepackage{subcaption}

\newtheorem{theorem}{Theorem}
\newtheorem{corollary}{Corollary}
\newtheorem{definition}{Definition}

\newtheorem{lemma}{Lemma}

\newtheorem{proposition}{Proposition}

\newcommand{\e}{{\mathbb{E}}}
\newcommand{\custmsz}{0.8}

\allowdisplaybreaks

\begin{document}

\title{The Hidden Perception Constraint in \\ Task-Aware Compression}

\author{Sahan Liyanaarachchi, Semih Akkoc \& Sennur Ulukus \\
University of Maryland \\
College Park, MD, USA \\
\texttt{\{sahanl,akkoc,ulukus\}@umd.edu}
\And
Aylin Yener \\
The Ohio State University \\
Columbus, OH, USA \\
\texttt{yener@ece.osu.edu}}

\maketitle

\begin{abstract}
    With the recent advancements of neural compressors, explicitly incorporating  perception constraints into the design of compression schemes has gained significant attention. Traditionally, these perception constraints ensure that the distribution of the reconstruction does not significantly deviate from the distribution of the source, thus attesting to the \emph{perceptual} quality of the reconstruction. In this work, we uncover several perception constraints that are naturally present in task-aware compression. In particular, we consider a problem where the primary task is reconstruction and the secondary task is classification (i.e., a statistical test). We study this problem at varying levels of domain information available to us and discuss how to utilize the naturally emerging perception constraints to design rate-minimal compression schemes that also maximize the utility of our secondary task. We show that in this setting, if the decision boundaries of the classifier are ill-defined (mismatch) for our source distribution, then matching onto a target distribution enhances our classification accuracy.
\end{abstract}

\section{Introduction}
Rate distortion theory has been a widely pursued interest among information theorists ever since its first formulation by Shannon \citep{shannon}. The rate-distortion (RD) problem involves finding the theoretical limit for compression such that the distortion between the reconstruction obtained from the compressed representation and the original source is within a pre-specified level. This constraint is formally known as the \emph{distortion} constraint and numerous variations of this problem have been studied  throughout literature by considering various distortion functions, availability of side information and for different network topologies \citep{wyner_side_info, cascaded_coding, many_decoders, yener_relay}. With the recent advancements of neural compressors, a new constraint known as the \emph{perception} constraint was identified in \citep{blau_rdp} as a necessary feature for modern compression schemes. While the distortion constraint ensures that our reconstruction is close to the original source, the perception constraint ensures that the distribution of the reconstruction is close to the source distribution. This results in more realistic reconstructions, which is a desired feature of modern compressors, but this comes at the expense of extra distortion.

This new problem of finding the optimal rate that simultaneously satisfies both the distortion and perception constraints is known as the rate-distortion-perception (RDP) problem and has gained significant attention recently. We refer to Section \ref{sec:rel_works} for the recent advances on this problem. In this work, we study a practically motivated variant of the above problem. In here, we explore how to compress while simultaneously maximizing the utility of a given task. In particular, we focus on the following problem: Suppose we want to transmit a compressed representation of our data to a remote server which uses the reconstructed data for a certain task. For instance, suppose that our source is a random variable that may originate from several hypotheses, $H_i$. We want to compress and transmit our observation of this source random variable $X$ and form a reconstruction $\hat{X}$ of it at the server (receiver) such that its distortion is within a pre-specified level. This is our primary goal. At the same time, suppose the server uses a generic or ill-defined statistical test (the test may be designed for a different data distribution or there may be a mismatch in prior distributions which was used to design the test) on the reconstructed data to identify the originating hypothesis. We want the error of this statistical test to be low as well. This is our secondary task.

In the context of machine learning, this can be viewed as using a pretrained classifier at the remote server to classify data coming from a distribution which is significantly different from the training distribution of the classifier. In here, the pretrained classifier can be considered as a third-party tool available at the sever which we use for our task. As a specific example, consider the presence (hypothesis $H_1$) and absence (hypothesis $H_0$) of a particular disease; and that the detector (classifier) has been trained with a data set that is dominated by a certain race; and that the new patient belongs to an underrepresented race. This paper tackles the problem of how to compress the data of the new patient with as few bits as possible, such that, the decompressed data at the detector ($\hat{X}$) has small distortion with the original data ($X$), and at the same time, the down-stream task of classification of the patient ($H_0$ or $H_1$) using the decompressed data has small probability of error. Essentially, we are trying to compress the data in a way to leverage a generic classifier to our advantage.

We tackle the above problem by considering varying levels of domain knowledge available to us. In particular, we identify the following three levels of information availability:
\begin{itemize}
     \item\textbf{Level 1:}   We know the distributions of the individual hypotheses of the source random variable and the decision regions of the classifier at the receiver side.
     
     \item\textbf{Level 2:} We know the individual distributions for each hypothesis of the source random variable but we do not know the exact decision regions at the receiver side. However, in this scenario, we assume that we do know that the classifier works well for some known distribution (i.e., yields low probability of error on a known distribution). In the computer vision domain, this can be viewed as a generic image classifier which is known to have a good inference accuracy on a common dataset such as ImageNet.
     
     \item\textbf{Level 3:} We do not know the individual distributions for each hypothesis of the source random variable and we do not know decision regions of the classifier at the receiver. Again, we assume that the classifier works well for a known distribution.
\end{itemize}
Under the above three levels of information availability, we find the optimal compression rates that satisfy the primary task while maximizing the performance of the secondary task which is classification. We identify that perception is a hidden constraint within this problem setup and formulate the above problem as an equivalent RDP problem. We show that when the exact decision regions of the classifier is not known, a good strategy is to compress the data so that the reconstruction's distribution is as close as possible to a target distribution for which we know the test yields good results.

\begin{figure}[t]
    \centering
    \includegraphics[width=\linewidth]{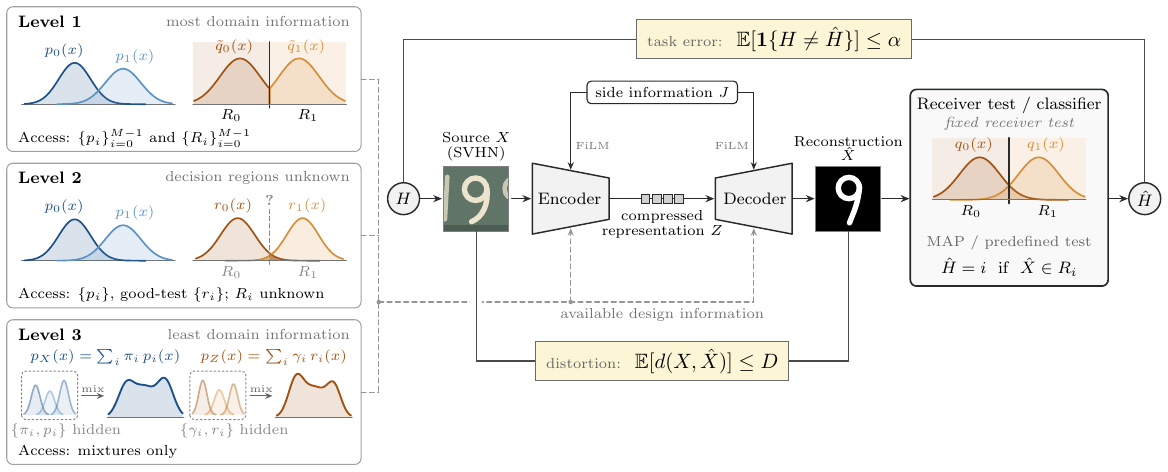}
    \caption{Diagram of the system model.}
    \label{fig:flow_diag}
\end{figure}

\section{Related Work}\label{sec:rel_works}
The RDP tradeoff was first formalized in \citep{blau_rdp}, which introduced the RDP function, where the authors show under mild regularity conditions such as convexity of the perception function, that any achievable rate must lie on the epigraph of the RDP function. The achievability of the above RDP function was later established in \citep{wagner_achievabilty} with the use of stochastic coding schemes in the presence of unlimited common randomness. With the use of soft covering lemma, this problem was later studied in \citep{wagner_finite} with finite common randomness and total variation as the perception function. Under the same setting, the achievable region of rates was fully characterized with the presence of side information in \citep{deniz_rdp_sideinfo}. The role of side information and the common randomness in the RDP tradeoff was then studied in \citep{wagner_gunduz_2026}, which separately treated perception under \emph{marginal realism} (marginal distribution of reconstruction should match the source distribution) and \emph{joint realism} (joint distribution of the side information and reconstruction should match the joint distribution of side information and source). The works in \citep{universal_rdp,vector_gauss_rdp} further studied the RDP problem. All of the above works investigate the problem of matching the distribution of the reconstruction on to the distribution of the original source.

In our work, we are interested in matching the reconstruction distribution on to a target distribution that is different from the source distribution. This notion of perception has been studied in several other occasions. In \citep{RDD}, this notion of perception was termed as \emph{deception} where they found the optimal compression rates when the reconstruction should exhibit statistical characteristics significantly different from the original distribution. On a separate work in \citep{yener_rdp_summarization}, the rate distortion framework for text summarization introduced in \citep{yener_rd_summarization} was extended by introducing a perception constraint which measures the perceptual quality between the summaries and a reference distribution instead of the original text distribution. 

In the domain of task-aware compression, there exist several works that fit our problem setting. A unified approach for incorporating classification accuracy into lossy image compression was introduced in \citep{Zhang_RDC_2023}, where they introduced classification-distortion-perception (CDP) function and the rate-distortion-classification (RDC) function; the former corresponds to minimizing the classification error for a given distortion and perception constraint. whereas the latter corresponds to minimizing the rate given the constraints on the distortion and classification accuracy. The distinction between the RDP, RDC and rate-perception-classification (RPC) functions was highlighted in \citep{Angela_2024} by deriving closed-form expressions of the above functions for Bernoulli and Gaussian source distributions. In \citep{Wang_RDPC_2025}, the authors extended the RDC function by incorporating a perception constraint and multiple classification constraints measured via the entropy of classification labels given the reconstruction. However, here too the perception constraint was measured with respect to the source distribution. In \citep{Nguyen_RDC_2025}, this entropy-based RDC function was studied in a setting with a universal encoder and multiple task-specific decoders. 
 
The closest to our work is the work in \citep{nguyen2026crossdomain}, which considers the problem of minimizing distortion with a rate and perception constraint while using an entropy-based classification constraint. This is an extension of the cross-domain lossy compression framework introduced in \citep{Liu_crossdomain_2022}. Here, they formulate the problem as an optimal transport problem where they consider the source distribution as a noisy version of a certain clean distribution. Therefore, perception in this case is measured between this noisy distribution and the clean distribution. They are interested in minimizing the expected distortion given a rate constraint, whereas we are interested in deriving the optimal compression rates for a given distortion in our setting. They solve the optimal transport problem by explicitly incorporating a perception constraint in addition to the entropy-based classification constraint, whereas we show that the perception constraint emerges naturally when realizing the constraint on classification accuracy, and that satisfying this naturally emerging perception constraint is sufficient to do well in terms of the classification accuracy.

\section{Problem Formulation}\label{sec:prob_form}
Suppose we have a total of $M$ hypotheses. Let $X$ be a random variable which is generated from hypothesis $H_i$ with probability $\pi_i$ following the distribution\footnote{We assume that the distributions of all continuous random variables  considered in this work are absolutely continuous with respect to the Lebesgue measure, and for discrete random variables the integrals over $\dd{x}$ must be interpreted as summations with respect to the corresponding probability mass function.} $p_i(x)$ and let $\mathcal{X}$ be the support of $X$. The realization of random variable $X$ is then compressed and sent to a remote server which reconstructs it as $\hat{X}\in \mathcal{X}$. At the remote server, this reconstruction will be subjected to a statistical test (classification) which was designed for a random variable $Y$ whose support is also $\mathcal{X}$, which originates from $H_i$ with probability $\sigma_i$ following the distribution $q_i(x)$. In other words we will be using a statistical test designed with incorrect domain knowledge to classify our reconstructed data. 

Suppose that $X$ came from the hypothesis $H$. Then, based on the reconstruction $\hat{X}$, we will be generating an estimate $\hat{H}$ of the original hypothesis $H$, using the available test at the server. In particular, let $R_i$ be the decision regions of the predefined test at the server with $R_i\cap R_j=\emptyset$ for $i\neq j$ and $ \bigcup_{i=0}^{M-1} R_i=\mathcal{X}$. Then, $\hat{H}=i$ if $\hat{X}\in R_i$. Let $p_e(\hat{X})=\e[\mathds{1}\{H\neq\hat{H}\}]$ be the probability of error of our test and let $d(X,\hat{X})$ denote an integrable distortion function which measures how close our reconstruction is to our original random variable. Our goal is to compress the random $X$ such that reconstruction $\hat{X}$ satisfies a constraint on the average distortion and the probability of error. In particular, we want our reconstruction to satisfy $\e[d(X,\hat{X)]}\leq D$ and $p_e(\hat{X})\leq\alpha$, where $D\in\mathds{R}_0^+$ and $\alpha\in[0,1]$ (see Fig.~\ref{fig:flow_diag}). Here, the distortion constraint ensures the semantic similarity between the original random variable and its reconstruction, and the constraint on the error probability ensures that we are simultaneously performing well at our task of interest which is to identify correctly the underlying hypothesis from which the random variable originated.

We will tackle the above problem at varying levels of domain knowledge that is available to us. To formally define the problem, we will start with a definition for achievability in this problem setting. In here, we will use $X^n$ to represent a collection $n$ i.i.d.~realization of $X$ and we use $X_{n,i}$ to represent the $i$th element of this collection.

\begin{definition}
    The tuple $(R,D,\alpha)$ is achievable if $\forall\epsilon>0$, $\exists n\in\mathbb{N}$, an encoding function $f:X^n\cross U\to \mathbb{N}$, and a decoding function $g:\mathbb{N}\cross U\to \hat{X}^n$, such that $K=f(X^n,U)$,  $\hat{X}^n=g(K,U)$ and satisfies,
    \begin{align}
        \frac{H(K|U)}{n}<R+\epsilon, \qquad 
        \frac{1}{n}\sum_{i=1}^n \e[d(X_{n,i},\hat{X}_{n,i})]\leq D, \qquad 
        \frac{1}{n}\sum_{i=1}^n p_e(\hat{X}_{n,i})\leq \alpha,
    \end{align}
    where $U$ is the common randomness independent of $X$ available to both the encoder and decoder.
\end{definition}

Next, to characterize the achievable region for a given distortion constraint $D$ and an error bound $\alpha$, we will define the rate-distortion-error (RDE) function $R(D,\alpha)$ as 
\begin{align}
    R(D,\alpha)= \inf_{p_{\hat{X}|X}}  I(X;\hat{X})
    \quad \text{such that}  \quad \e[d(X,\hat{X})]\leq D, \quad p_e(\hat{X})\leq\alpha.
    \label{eqn:main}
\end{align}
 Note that this function is not well-defined for every $(D,\alpha)$ pair. This is because the decision regions at the server are suboptimal. For instance, suppose $D=0$ and we use the mean squared error (MSE) as the distortion metric, then even with infinite rate, we may not be able to achieve the desired error probability since $R_i$s are not optimal decision regions for $X$. Therefore, we will define this function only if $S_{D,\alpha}\neq\emptyset$, where $ S_{D,\alpha}=\{p_{\hat{X}|X}: \ \e[d(X,\hat{X})]\leq D, \ p_e(\hat{X})\leq\alpha\}.$ However, Proposition \ref{thrm:achieve_reg} shows whenever a tuple $(R,D,\alpha)$ is achievable, the function $R(D,\alpha)$ is well-defined.

\begin{proposition}\label{thrm:achieve_reg}
    The tuple $(R,D,\alpha)$ is achievable iff $S_{D,\alpha}\neq \emptyset$ and $R\geq R(D,\alpha)$.
\end{proposition}

The proof of Proposition \ref{thrm:achieve_reg} is given in Appendix \ref{appen:thrm_achieve_reg}. Proposition \ref{thrm:achieve_reg} simply states that $R(D,\alpha)$ defines the minimum achievable compression rate satisfying a given distortion constraint $D$ and an error bound $\alpha$. Therefore, finding the minimum compression rate is equivalent to solving the RDE function over the set of stochastic mappings $p_{\hat{X}|X}$ from $X$ to $\hat{X}$ that are feasible under the given constraints. Now, given a stochastic mapping $p_{\hat{X}|X}$, the probability of error is 
\begin{align}
    p_e(\hat{X})=\sum_{i=0}^{M-1}\int_{\mathcal{X}}\pi_ip_i(x)\int_{\bar{R}_{i}}p_{\hat{x}|x}\dd{\hat{x}} \dd{x} =1-\sum_{i=0}^{M-1}\int_{\mathcal{X}}\pi_ip_i(x)\int_{R_{i}}p_{\hat{x}|x}\dd{\hat{x}} \dd{x},\label{eqn:pe_hatx}
\end{align}
where $\bar{R}_i=\mathcal{X}\setminus R_i$.

Let $R_i^*$ be the decision regions of the optimal hypothesis test constructed for $X$ where  $R_i^*\cap R_j^*=\emptyset$ for $i\neq j$ and $ \bigcup_{i=0}^{M-1} R_i^*=\mathcal{X}$. Let $p_e^*$ be the error probability of the optimal hypothesis test. Then, in Lemma~\ref{lem:probs_ineq}, whose proof is given in Appendix \ref{appen:lem_probs_ineq}, we show that $p_e(\hat{X})\geq p_e^*$. Hence, the above problem is feasible only if $\alpha \geq p_e^*$. 

\begin{lemma}\label{lem:probs_ineq}
    For any stochastic mapping $p_{\hat{X}|X}$, $p_e(\hat{X})\geq p_e^*$ with equality iff almost everywhere,
    \begin{align}
        \int_{R_i}p_{\hat{x}|x}\,d\hat{x}=1,~\text{if}~x\in R_i^*.\label{eqn:opt_cond1}
    \end{align}
\end{lemma}

Therefore, we will only consider values of $\alpha\geq p_e^*$. In the next section, we will focus on the setting where $\alpha=p_e^*$, and show that the above problem can be also formulated as an RDP problem.

\section{Compression with Optimal error probability}
In this section, we will study the above problem under the special case where $\alpha=p_e^*$. We will call this setup the rate-distortion-minimum-error (RDME) problem.  Therefore, our stochastic mappings must satisfy \eqref{eqn:opt_cond1}. Let us denote a stochastic map that satisfies these conditions by $p^*_{\hat{X}|X}$. Let $J$ be a discrete random variable which assumes $J=i$ if $X\in R_i^*$. Let $p(x)=\sum_{i=0}^{M-1}\pi_ip_i(x)$ and $P_i=\int_{R_i^*}p(x)\,dx$. Then, $J=i$ with probability $P_i$. Here, $J$ will denote the estimate of $H$ if we carryout the optimal hypothesis test on $X$. Note that, $X$ can be also represented using $J$ as $X=X_J$
where $X_i\sim \frac{1}{P_i}p(x)\mathds{1}\{x\in R_i^*\}$. Let $\hat{X}_i$ be the reconstruction generated when $J=i$. Now, under $p^*_{\hat{X}|X}$, we have that $J$ is fully deterministic given either $\hat{X}$ or $X$. Therefore,
\begin{align}
    I(X;\hat{X})=I(X;\hat{X})+I(X;J|\hat{X})=I(X;J)+I(X;\hat{X}|J)=H(J)+\sum_{i=0}^{M-1}P_iI(X_i;\hat{X}_i).
\end{align}
Let $S_D=\{D:  S_{D,p_e^*}\neq \emptyset\}$. Then, for $D\in S_D$, the optimal compression rate $R(D)$ will be given by $R(D)=H(J)+R'(D)$, where
\begin{align}
    R'(D)=\inf_{p^*_{\hat{X}|X}} \sum_{i=0}^{M-1}P_iI(X_i;\hat{X}_i) \quad 
     \text{such that} \quad \sum_{i=0}^{M-1}P_i\e[d(X_i,\hat{X}_i)]\leq D. \label{eqn:opt_err_rate_og}
\end{align}
Therefore, in order to achieve the optimal error probability, our compression scheme must always send at least the bits necessary to represent our own estimate of the hypothesis (i.e., $J$) done at the transmitter side on $X$. 

Next, note that, we can also represent the above optimization problem as an RDP problem as  follows,
\begin{align}
    R'_P(D)= \inf_{p_{\hat{X}|X}}  \sum_{i=0}^{M-1}P_iI(X_i;\hat{X}_i)
    \ \text{such that} \ \sum_{i=0}^{M-1}P_i\e[d(X_i,\hat{X}_i)]\leq D, \ \sum_{i=0}^{M-1}P_i D(P_{\hat{X}_i}||P_{Y_i})\leq P,
     \label{eqn:tJprob}
\end{align}
where $P_{\hat{X}_i}$ is the distribution of $\hat{X}_i$ and $P_{Y_i}(x)=\tilde{q}_i(x)=\frac{1}{Q_i}q(x)\mathds{1}\{x\in R_i\}$  with $q(x)=\sum_{i=0}^{M-1}\sigma_iq_i(x)$, $Q_i=\int_{R_i}q(x)\,dx$,  and $P<\infty$ being our perception constraint. Note that, we have changed $p^*_{\hat{X}|X}\to p_{\hat{X}|X}$ since the perception constraint ensures that any feasible stochastic mapping for the above problem must satisfy \eqref{eqn:opt_cond1}. Also, note that, as $P\to\infty$, the above optimization problem will converge to the original optimization problem for the RDME problem. Moreover, in this scenario, we have that $R_P(D)\geq R(D)$, where $R_P(D)=H(J)+R'_P(D)$.

\section{Compression via Distribution Matching}
In Section \ref{sec:prob_form}, we formally defined the  RDE function and in the previous section, we noticed how the RDME problem can be formulated as an RDP problem. However, both of these problem formulations required knowledge of the decision regions at the receiver (Level 1). These decision regions depend both on the exact distributions it was designed for and the particular test that is used (e.g., Neyman-Pearson or MAP). In a more general setting, if the task at hand is something more complex, such as image classification, these decision regions depend on both the structure of the classifier at the receiver as well as its training distribution. In this section, we discuss compression schemes designed without explicit knowledge about the decision regions of the test at the receiver. However, in this scenario, we assume that we know the test works well (low error probability) for some known distribution (Level 2). Let this test distribution be defined as follows: the input $x$ to the statistical test is generated from $H_i$ with probability $\gamma_i$ following the distribution $r_i(x)$. Now, Theorem \ref{thrm:lvl2_err_bound} gives an important relationship between $p_e(\hat{X)}$ and the distributions $r_i(x)$.

\begin{theorem}\label{thrm:lvl2_err_bound}
  Let $r_e$ denote the probability of error for a known test distribution $r_i(x)$ when using the statistical test at the receiver. Then, the following error bound is satisfied
  \begin{align}
      p_e^*\leq p_e(\hat{X})\leq p_e^*+Cr_e+\sum_{i=0}^{M-1}P_i\|P_{\hat{X}_i}-r_i\|_{TV}
  \end{align}
  where $C=\max_i\frac{P_i}{\gamma_i}$ and $\|.\|_{TV}$ is the total variation distance.
\end{theorem}

\begin{corollary}\label{cor:lvl2_err_bound}
    The following error bound is satisfied, 
    \begin{align}
    p_e^*\leq p_e(\hat{X})\leq p_e^*+Cr_e+\sqrt{\frac{\hat{P}}{2}
      }
    \end{align}
    where $\hat{P}=\sum_{i=0}^{M-1}P_i D(P_{\hat{X}_i}||r_i)$.
\end{corollary}

The proofs of Theorem \ref{thrm:lvl2_err_bound} and Corollary \ref{cor:lvl2_err_bound} are given in Appendix \ref{appen:thrm_lvl2_err_bound}. In order to explicitly represent the probability of error, we needed to know the decision regions of the test at the receiver. Corollary \ref{cor:lvl2_err_bound} presents us with an alternative way to formulate the constraint on the probability of error. In particular, we can represent the optimal compression problem in \eqref{eqn:main} as,
\begin{align}
    R^{(2)}(D,\alpha)= \inf_{p_{\hat{X}|X}}  I(X;\hat{X}) 
    \quad \text{such that} \quad \e[d(X,\hat{X})]\leq D, \quad  \sum_{i=0}^{M-1}P_i D(P_{\hat{X}_i}||r_i)\leq P_{\alpha},
    \label{eqn:RDP_2}
\end{align}
where $P_{\alpha}$ is a function of $\alpha$, e.g., $P_\alpha=2(\alpha-p_e^*-Cr_e)^2$ if $\alpha\geq p_e^*+Cr_e$. However, since the perception constraint only forms an upper bound on the probability of error, we have that $R(D,\alpha)\leq R^{(2)}(D,\alpha)$ even when both of the problems are feasible. We further note that, if we are able to send $J$ to the receiver side, then the optimal compression problem becomes $R'_P(D)$ by replacing $P_{Y_i}$ with $r_i$. The above formulations represents Level 2 of our information availability.

Next, we look at the more general setting where we do not have any information about the individual distributions of the source random variable (Level 3). Here, we will only have access to the mixture distribution ($p(x)$ of $X$). In such a setting, one may try to formulate the optimal compression problem as follows,
\begin{align}
    R^{(3)}(D,\alpha)= \inf_{p_{\hat{X}|X}}  I(X;\hat{X}) \quad
    \text{such that} \quad \e[d(X,\hat{X})]\leq D, \quad  D(P_{\hat{X}}||P_Z)\leq P,
    \label{eqn:RDP_3}
\end{align}
where $P_{Z}(x)=\sum_{i=0}^{M-1}\gamma_ir_i(x)$. Here, we would want $P$ to be as small as possible so that we are closer to a sample from the well-known distribution. Our experimental results suggest that the viability of this approach mainly depends on how informative the distortion measure is with respect to the source observation and how informative the mixture $r(x)$ is with respect to the decision regions of the classifier.

\section{Numerical Results} \label{sec:num_res}
In this section, we present numerical results on  discrete alphabets for a binary hypothesis test ($M=2)$. The exact derivations of our rate function for Bernoulli sources can be found in  Appendix \ref{appen:Bernouli_1}. In here, we consider a quaternary alphabet with a binary hypothesis test and evaluate how our rate and error probability vary at different distortion levels measured using the Hamming distortion. For these experiments, we set $\mathcal{X}=\{0,1,2,3\}$ and the prior probabilities to $\pi_0=\sigma_0=0.5$.  The optimal decision regions $R_i^*$ at the transmitter side is obtained through MAP rule and we consider that the test at the receiver is also MAP but its decision regions $R_i$ are obtained following the distributions of $q_i(x)$s. The distributions $p_i(x)$ and $q_i(x)$ for $i=0,1$ along with the corresponding decision regions are presented analytically in Appendix \ref{appen:xtra_discrete}.

\begin{figure*}[t]
    \centering

    \begin{minipage}[t]{0.3\textwidth}
        \vspace{0pt}
        \centering
        \includegraphics[width=\linewidth]
            {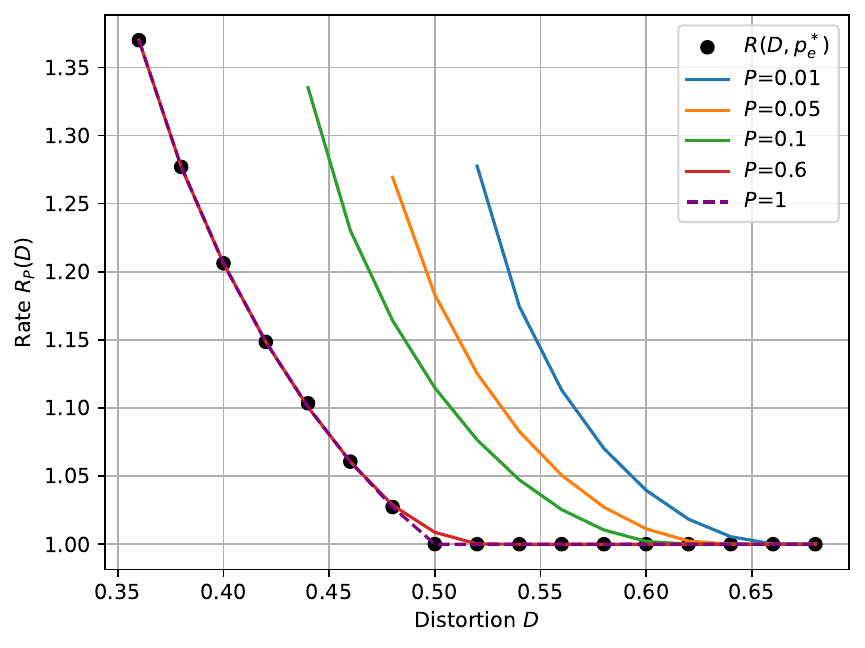}
        \caption{Variation of
            $R_P(D)$ vs distortion 
            with varying perception levels $P$
            for the RDME problem.}
        \label{fig:Lvl1_R_vs_D}
    \end{minipage}\hfill
    \begin{minipage}[t]{0.66\textwidth}
        \vspace{0pt}
        \centering
        \begin{subfigure}[t]{0.45\linewidth}
            \vspace{0pt}
            \centering
            \includegraphics[width=\linewidth]
                {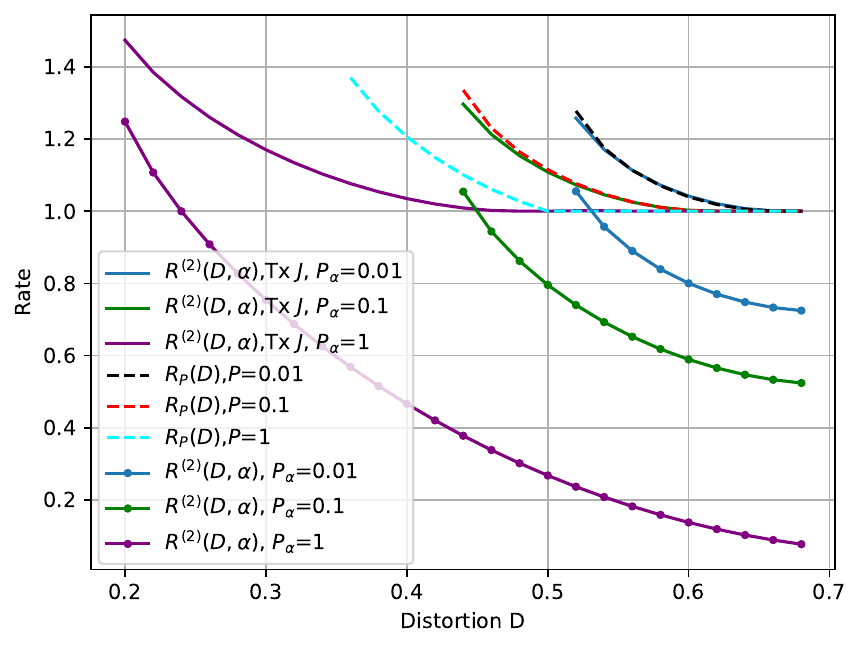}
            \caption{rate vs distortion}
            \label{fig:L2_R_vs_D}
        \end{subfigure}~
        \begin{subfigure}[t]{0.45\linewidth}
            \vspace{0pt}
            \centering
            \includegraphics[width=\linewidth]
                {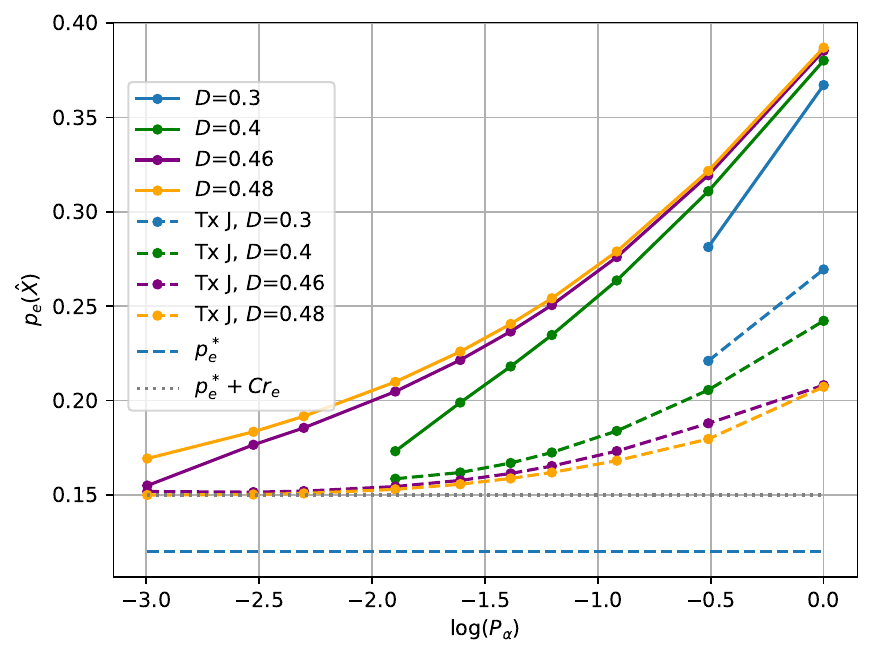}
            \caption{$p_e(\hat{X})$ vs perception}
            \label{fig:L2_pe_vs_P}
        \end{subfigure}
        \caption{Variation of rate and $p_e(\hat{X})$
            when transmitting side information $J$
            separately versus not transmitting $J$.}
    \end{minipage}

    \par\vspace{1.5em}

    \begin{minipage}[t]{0.81\textwidth}
        \vspace{0pt}
        \centering
        \begin{subfigure}[t]{0.39\linewidth}
            \vspace{0pt}
            \centering
            \includegraphics[width=\linewidth]
                {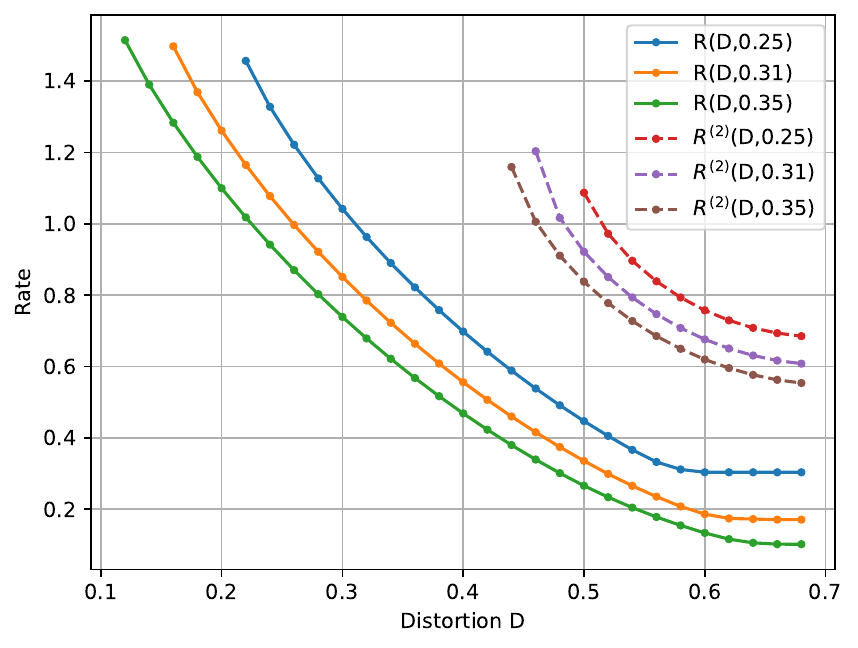}
            \caption{rate}
            \label{fig:L1_vs_L2_rate}
        \end{subfigure}~
        \begin{subfigure}[t]{0.39\linewidth}
            \vspace{0pt}
            \centering
            \includegraphics[width=\linewidth]
                {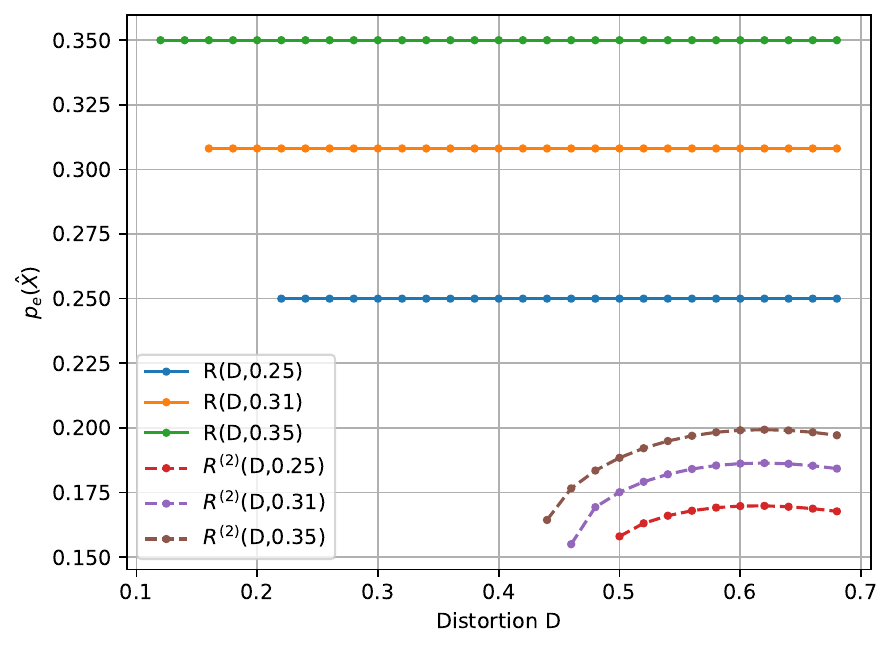}
            \caption{$p_e(\hat{X})$}
            \label{fig:L1_vs_L2_pe}
        \end{subfigure}
        \caption{Variation of rate and $p_e(\hat{X})$
            vs distortion at information Levels 1 and 2.}
    \end{minipage}
\end{figure*}

In the first experiment, we consider information Level 1 and look at the RDME problem. Here, we consider how the rate varies with distortion at different perception levels enforced in \eqref{eqn:tJprob}. As illustrated in Fig.~\ref{fig:Lvl1_R_vs_D}, as we increase the perception constraint $P$, the rate curves tend to converge and finally settle on the rate curve obtained by directly solving \eqref{eqn:opt_err_rate_og} as claimed. In the next experiment we operate at Level 2 and consider the following two scenarios: (i) We transmit the side information $J$ losslessly and try to minimize the additional bits required for the specific reconstruction. (ii) We do not transmit $J$ explicitly, but instead try to minimize the overall rate. For these two scenarios, we examine how the rate varies with distortion for different perception levels and how probability of error varies with perception at different distortion levels. Here, for simplicity, we use $r_i(x)=q_i(x)$ and $\gamma_i=\sigma_i$ for $i=0,1$. This essentially means that we are matching to the training distribution of the test at the receiver. However, this does not mean that we know the decision regions at the receiver since the test could be different (e.g., Neyman-Pearson test instead of MAP). As depicted in Fig.~\ref{fig:L2_R_vs_D}, transmitting $J$ losslessly induces higher rates for a given perception and distortion level. However, as illustrated in Fig.~\ref{fig:L2_pe_vs_P}, transmitting $J$ lowers the error probability as well. Moreover, it can be observed that for a given distortion level, as we lower the perception constraint, the probability of error approaches the limit of our bound (small $\hat{P}$) in Corollary \ref{cor:lvl2_err_bound}. This shows that for a given distortion level, it is better to operate at the tightest possible perception level that is feasible to minimize the classification error.

Next, we evaluate how rate and probability of error vary when we operate at Level 1 and Level 2. Here, the relationship between $P_\alpha$ and $\alpha$ is defined using $P_\alpha=2(\alpha-p_e^*-Cr_e)^2$. As shown in Fig.~\ref{fig:L1_vs_L2_rate} and Fig.~\ref{fig:L1_vs_L2_pe}, when we operate at Level 1, we obtain lower rates compared to Level 2. However, at Level 1, we observe that this lower rate is mostly achieved when the constraint on the probability of error is satisfied with equality thus resulting in higher error whereas in Level 2, we always observe that the obtained error probability is much smaller compared to that of Level 1. This is because at Level 2, we are using an upper bound on error probability to minimize the rate. Though this limits the feasible distortion levels, this is a viable approach when the exact decision regions are unknown. Additionally, it is interesting to see that the best error rate obtained in Level 2 is attained when the distortion constraint is not too tight so as to make the problem infeasible and when it is not too relaxed so as not to convey any useful information.

\section{Simulations on Neural Classifiers}
\label{sec:img_exp}
In this section, we move from discrete alphabets to a setting where the test at the receiver is a neural classifier and the decision regions $R_i$ are not available in closed-form. This is precisely the situation described in Section~\ref{sec:prob_form} for which Level 2 was introduced: We do not know $R_i$, but we know a distribution on which the test performs well.

We look at the digit classification problem and consider $M=10$ hypotheses corresponding to the ten digit classes. The source $X$ is drawn from SVHN dataset~\citep{svhn} and the distribution $r_i(x)$ for which the receiver's test was designed is the distribution of MNIST images~\citep{mnist} belonging to class $i$, both rendered as $32\times32$ RGB images so that the two share common support $\mathcal{X}$. We set $\pi_i=\sigma_i=\gamma_i=1/M$ for all $i$, so that the domain mismatch is entirely a mismatch of the distributions within each class and not of the priors. The test at the receiver is a ResNet-9 classifier trained only on MNIST, its error on its own domain is $r_e=0.0077$. Since the transmitter has no access to this classifier, the decision regions $R_i$ are unknown to the compression scheme, and neither the classifier nor its gradients are ever used while designing the encoder and decoder. The estimate $J$ of the hypothesis is produced at the transmitter by a second ResNet-9 classifier trained only on SVHN, whose accuracy is $0.921$. This classifier plays the role of the optimal test at the transmitter and yields $p_e^*=0.0790$ and $C=1.0867$. Details of both classifiers and of the compression architecture are given in Appendix \ref{appen:img_exp_params}.

Two features of this setting are worth emphasizing before looking at the results. First, the mismatch is severe: Applying the receiver's test directly to $X$ yields an error probability of $0.8930$, which is close to $1-1/M=0.9$. The decision regions of the receiver are therefore bad for the source, which is the regime in which Theorem \ref{thrm:lvl2_err_bound} predicts that matching onto $r_i$ should help. Second, unlike the discrete examples of Section \ref{sec:prob_form}, here we cannot solve \eqref{eqn:RDP_2} directly. We therefore adopt the standard approach of \citep{blau_rdp} and trace the achievable region rather than impose a constraint: We fix the rate by construction and sweep the Lagrangian
\begin{align}
    \min_{\text{enc},\text{dec}} \quad \sum_{i=0}^{M-1}P_i\e[d(X_i,\hat{X}_i)] + \lambda\sum_{i=0}^{M-1}P_i\,\delta\!\left(P_{\hat{X}_i},P_{Y_i}\right),
    \label{eqn:img_lagrangian}
\end{align}
where $\delta(\cdot,\cdot)$ is a divergence estimated adversarially and $\lambda\ge0$ selects an operating point on the distortion-perception tradeoff. The rate is fixed rather than penalized because the bottleneck is a vector quantizer with $L$ tokens over a codebook of $K$ codewords, which gives exactly $R=L\log_2K$ bits per image irrespective of the weights. The side information $J$ costs $H(J)=\log_2M$ bits when it is transmitted. Sweeping $\lambda$ at each $(L,K)$ therefore produces a family of achievable $(R,D,P,p_e)$ tuples rather than a single constrained solution, and every point in the plots below is an operating point that was actually realized by a trained encoder-decoder pair. In this section, we opt to use the Wasserstein $1$-distance $(W_1)$ instead of the total variation (TV) distance to quantify perception. This is because the $W_1$ distance can be expressed as a functional optimization problem using the Kantorovich-Rubenstein duality leading to the WGAN architecture which is well-known to work well for image generation tasks. Even though TV distance can be equivalently expressed as a functional optimization problem in the space of bounded functions, we see that the smoothness constraint enforced in the Kantrovich duality of the $W_1$ distance is necessary to generate good reconstructions from the target distribution. 

\begin{figure}[t!]
    \centering
    \begin{subfigure}[b]{0.33\textwidth}
        \centering
        \includegraphics[width=\custmsz\textwidth]{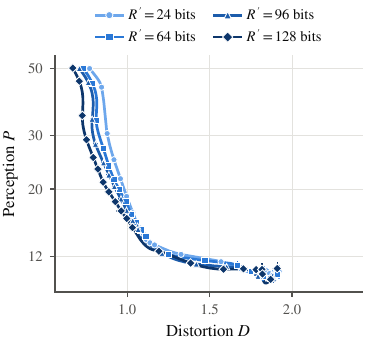}
        \caption{perception vs distortion}
        \label{fig:img_L2J_PvD}
    \end{subfigure}%
    ~ 
    \begin{subfigure}[b]{0.33\textwidth}
        \centering
        \includegraphics[width=\custmsz\textwidth]{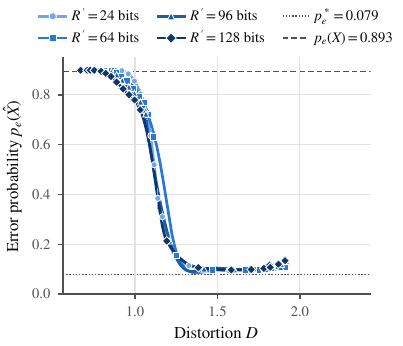}
        \caption{error vs distortion}
        \label{fig:img_L2J_EvD}
    \end{subfigure}%
    ~ 
    \begin{subfigure}[b]{0.33\textwidth}
        \centering
        \includegraphics[width=\custmsz\textwidth]{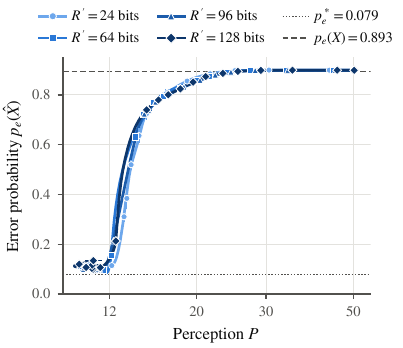}
        \caption{error vs perception}
        \label{fig:img_L2J_EvP}
    \end{subfigure}%
    \caption{Variation of distortion, perception and classification error at different rates (true rate $R=R'+H(J)$) for Level 2 with lossless transmission of $J$. }
    \label{fig:img_L2J}
\end{figure}

\begin{figure}[t!]
    \centering
    \begin{subfigure}[b]{0.33\textwidth}
        \centering
        \includegraphics[width=\custmsz\textwidth]{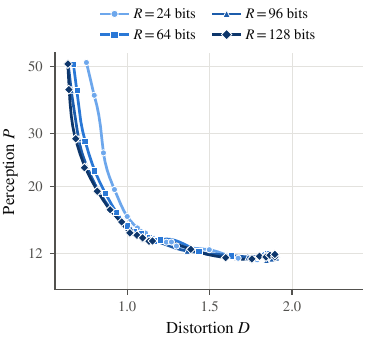}
        \caption{perception vs distortion}
        \label{fig:img_L2NoJ_PvD}
    \end{subfigure}%
    ~ 
    \begin{subfigure}[b]{0.33\textwidth}
        \centering
        \includegraphics[width=\custmsz\textwidth]{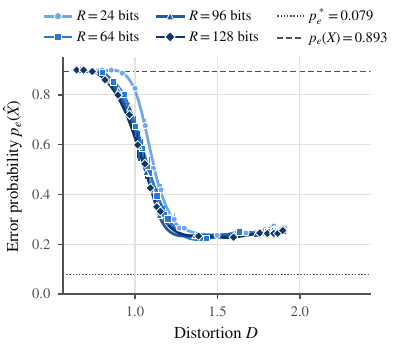}
        \caption{error vs distortion}
        \label{fig:img_L2NoJ_EvD}
    \end{subfigure}%
    ~ 
    \begin{subfigure}[b]{0.33\textwidth}
        \centering
        \includegraphics[width=\custmsz\textwidth]{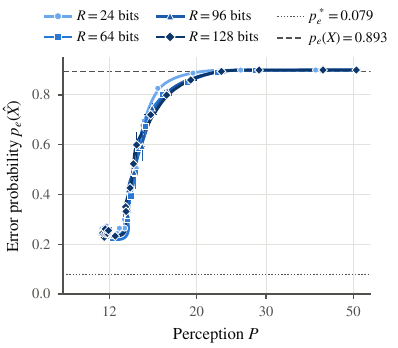}
        \caption{error vs perception}
        \label{fig:img_L2NoJ_EvP}
    \end{subfigure}%
    \caption{Variation of distortion, perception and classification error at different rates for Level 2 without explicitly transmitting $J$.}
    \label{fig:img_L2NoJ}
\end{figure}

To evaluate the interplay between distortion, perception and classification error, we consider the two configurations for Level 2 described in Section \ref{sec:prob_form}, where in the first configuration we send $J$ losslessly and in the second we do not. As illustrated in Fig.~\ref{fig:img_L2J} and Fig.~\ref{fig:img_L2NoJ}, when the rate increases our rate distortion curves tend to move towards the low distortion and high perception regime for both configurations. Further, for both configurations, we see that the classification error decreases significantly when the distortion constraint is not too tight and when the perception constraint is binding. This aligns with our theoretical insights and further  validates our claim that significant gains in terms of classification accuracy can be obtained when matching on to a target distribution rather than the source distribution in this setting. Moreover, we also see that transmitting $J$ lossessly further enhances our classifications accuracy at the cost of only $H(J)\approx 3.32$ bits.

\begin{figure}[t!]
    \centering
    \begin{subfigure}[b]{0.33\textwidth}
        \centering
        \includegraphics[width=\custmsz\textwidth]{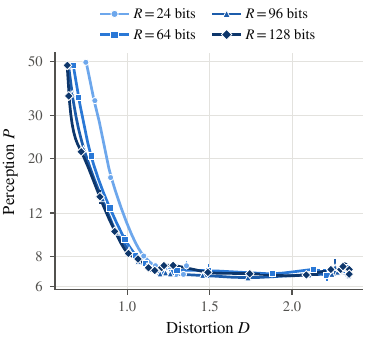}
        \caption{perception vs distortion}
        \label{fig:img_L3_PvD}
    \end{subfigure}%
    ~ 
    \begin{subfigure}[b]{0.33\textwidth}
        \centering
        \includegraphics[width=\custmsz\textwidth]{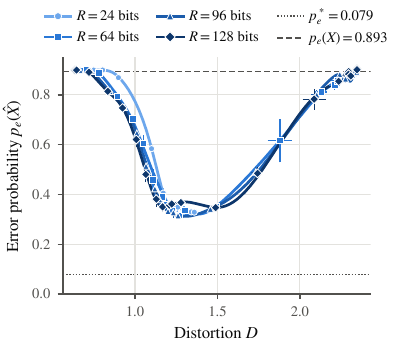}
        \caption{error vs distortion}
        \label{fig:img_L3_EvD}
    \end{subfigure}%
    \caption{Variation of distortion, perception and classification error at different rates for Level 3.}
    \label{fig:img_L3}
\end{figure}

Next, we consider the setting described in Level 3 where we try to match on to the mixture distribution. For this scenario, we see an interesting phenomenon: In Fig.~\ref{fig:img_L3}, when the distortion constraint is not too tight, we see that lower perception values lead to lower classification error. However, when the distortion constraint becomes too relaxed, lower perception yields higher classification errors. This is due to the fact that when the distortion constraint is very relaxed, it does not convey any useful information. Therefore, the reconstruction is generated from the target distribution independent of the source image. The reconstructions for Levels 2 and 3 are presented in Appendix \ref{appen:xtra_results}. 

\section{Conclusions}
In this work, we investigated rate-minimal compression schemes that simultaneously enable us to leverage an ill-trained classifier to our advantage. We showed that perception constraints naturally emerge out of the classification constraints defined in the problem. Through theory and extensive simulations, we showed that under this setting, matching on to a target distribution enhances classification accuracy as opposed to reconstructing similar to the source distribution.

\subsection*{AI use statement}
LLMs were used for literature review and as a programming assistant for implementing experimental code according to specifications and methodological choices provided by the authors. All AI-assisted outputs were reviewed and verified by the authors. LLMs were not used to derive or verify the proofs, generate synthetic datasets, or determine the scientific conclusions of the work. We take responsibility for the final content of this work, including text, claims or artifacts produced with the aid of generative AI.

\subsection*{Reproducibility statement}
Code, configurations and scripts that generate figures are available at \url{https://anonymous.4open.science/r/Task-Aware-Compression}.

\textbf{Theory.} All assumptions and the complete proofs which are in Appendixes are self-contained and no result relies on computation.

\textbf{Discrete experiments (Section~\ref{sec:num_res}).}
The RDP functions are obtained by solving convex programs directly on the quaternary and binary alphabets, with \texttt{scipy.optimize} over a grid of $(D,P)$ pairs. The three problem builders differ only in the constraints. The source and test distributions and both sets of decision regions are given in Table~\ref{tab:quat_dist}, the priors are $\pi_0=\sigma_0=0.5$, and the distortion is the Hamming distance. We release the solver together with the Streamlit interface used to produce the figures, which shows the distributions, the distortion, the $(D,P)$ grid and the perception levels as controls and plots the curves.

\textbf{Image experiments (Section~\ref{sec:img_exp}).}
Both datasets are the standard \texttt{torchvision} distributions of SVHN and MNIST. Each run writes the achieved operating point of every trained config as a CSV rate, distortion, perception, $p_e(\hat X)$, the two components of the distortion, and the constraints Section~\ref{sec:img_exp} is reproduced by:

\begin{verbatim}
python experiments.py --mode rate-sweep \
    --family digits_rgb --tx svhn --rx mnist \
    --resolution 32 --uniform-priors \
    --arch resnet9 --clf-epochs 100 \
    --clf-batch 512 --clf-lr 0.002 \
    --perception gan --distortion mse+feature \
    --feature-weight 1.0 \
    --feature-source transmitter_classifier \
    --feature-layer deep \
    --lk-pairs "(8,32),(8,64),(16,16),(16,64),(32,16)" \
    --lambdas "0,1,5,10,20,35,50,75,90,100,200,300" \
    --lambda-scaling balanced \
    --latent-dim 128 --n-down 2 --epochs-per-point 300 \
    --batch 800 --eval-batch 800 --eval-critic-steps 300\
    --decoder-noise 0.5 --seeds 4 \
    --configs sendJ_cond_PYj,sendJ_cond_PXj,noJ_cond_PYj,noJ_cond_PXj \
    --save-models --out results/
\end{verbatim}

with the Level 3 configuration \texttt{noJ\_uncond\_PY} run separately. Every point is checkpointed as it completes, so the sweep resumes rather than restarts, and the two classifiers are trained once and reused across all configurations from a cache keyed by their configuration hash. The runs reported here used NVIDIA A100 GPU and took approximately $80$ GPU-hours in total.

\textbf{What is and is not deterministic.}  
The data splits, the class sampling and all weight initializations are seeded (\texttt{split\_seed} $=0$, four compressor seeds per operating point), and the classifier cache makes $g_X$ and $g_Y$ same across configurations. First, GPU output being non-deterministic means a rerun on different hardware will not reproduce the numbers exactly. We report the mean over four seeds and the figures show the mean points. Second, the perception term is an adversarial estimate and is therefore non-stationary. It depends on how well the critic has been trained, which is why the critic is trained at every step including $\lambda=0$ and refit for a further 300 steps with the frozen weights before the value is recorded.

\bibliographystyle{iclr2027_conference}
\bibliography{refs}

\appendix
\section{Proof of Proposition \ref{thrm:achieve_reg}}\label{appen:thrm_achieve_reg}
The converse and achievability proofs follow similar to that of \citep{RDD}. We have used common randomness in this setup to make the achievability proof simpler and to facilitate  the use of stochastic encoders and decoders when perception constraints are involved.

\subsection{Converse}
From \eqref{eqn:pe_hatx}, we have that the $p_e(\hat{X})$ is linear with respect $p_{\hat{X}|X}$. Further, we have that $I(\hat{X};X)$ is convex and $\e[d(X,\hat{X})]$ is linear with respect to $p_{\hat{X}|X}$. Hence, the set $S=\{(D,\alpha):S_{D,\alpha}\neq \emptyset\}$ is convex and $R(D,\alpha)$ is convex on $S$. Additionally, the monotonicity of $R(D,\alpha)$ follows from the fact $S_{D,\alpha}\subseteq S_{D',\alpha '}$ for $D\leq D'$ and $\alpha\leq\alpha'$. Now, suppose $(R,D,\alpha)$ is achievable and fix $\epsilon>0$. Therefore,  $\exists n$ and functions $f(\cdot)$ and $g(\cdot)$ such that,
\begin{align}
    n(R+\epsilon)&> H(K|U)\geq I(X^n;K|U)=I(X^n;K,U)\\
    & \geq I(X^n;\hat{X}^n)\geq \sum_{i=1}^{n}I(X_{n,i};\hat{X}^n|X^{i-1})\\
    &=\sum_{i=1}^nI(X_{n,i};\hat{X}^n,X^{i-1})\geq \sum_{i=1}^nI(X_{n,i};\hat{X}_{n,i})
\end{align}
We can construct single-letter stochastic mappings that achieve the rate $I(X_{n,i};\hat{X}_{n,i})$  with distortion $D_i=\e[d(X_{n,i},\hat{X}_{n,i})]$ and probability of error $p_e^{(i)}=p_e(\hat{X}_{n,i})$, by isolating only the $i$th term of the above encoder-decoder scheme defined using $f(\cdot)$ and $g(\cdot)$. Therefore, based on the convexity of $S$, we have that $(\tilde{D},\tilde{\alpha})\in S$, where $\tilde{D}=\frac{1}{n}\sum_{i=1}^{n}D_i$ and $\tilde{\alpha}=\frac{1}{n}\sum_{i=1}^np_e^{(i)}$. Hence, $S_{\tilde{D},\tilde{\alpha}}\neq \emptyset$. Further, since $\tilde{D}\leq D$ and $\tilde{\alpha}\leq \alpha$, it follows that $S_{\tilde{D},\tilde{\alpha}}\subseteq S_{D,\alpha}$ and thereby $S_{D,\alpha} \neq \emptyset$. Now, from the convexity of $R(D,\alpha)$, we have that
\begin{align}
    n(R+\epsilon)&>\sum_{i=1}^nI(X_{n,i};\hat{X_{n,i}})\geq n\sum_{i=1}^{n}\frac{1}{n}R(D_i,p_e^{(i)})\geq nR(\tilde{D},\tilde{\alpha})\geq nR(D,\alpha).
\end{align}
Since this is true for any $\epsilon>0$, we have $R\geq R(D,\alpha)$.

\subsection{Achievability}
The achievability follows from the strong function representation lemma (SFRL) \citep{SFRL}. Suppose $R(D,\alpha)$ is well-defined and let $\epsilon>0$. Then, there exists a mapping $p_{\hat{X}|X}$ and thereby a $\hat{X}$ such that the distortion and probability of error constraints are satisfied with $I(X;\hat{X})< R(D,\alpha)+\frac{\epsilon}{2}$. Now, from SFRL there exists a random variable $U$ independent of $X^n$ and functions $f(\cdot)$ and $g(\cdot)$ such that $K=f(X^n,U)$, $\hat{X}^n=g(K,U)$ and 
\begin{align}
    H(K|U)&< I(X^n;\hat{X}^n)+\log(I(X^n;\hat{X}^n)+1)+5\\
           &= n I(X;\hat{X})+\log(nI(X;\hat{X})+1)+5. 
\end{align}
Therefore, by selecting $n$ large enough such that $\frac{\log(nI(X;\hat{X})+1)+5}{n}\leq \frac{\epsilon}{2}$,
we have that
\begin{align}
    \frac{H(K|U)}{n}<I(X;\hat{X})+\frac{\epsilon}{2}<R(D,\alpha)+\epsilon.
\end{align}
Therefore, any $(R,D,\alpha)$ is achievable for any $R\geq R(D,\alpha)$.

\section{Proof of Lemma \ref{lem:probs_ineq}}\label{appen:lem_probs_ineq}
Using \eqref{eqn:pe_hatx}, we have that
\begin{align}
    p_e(\hat{X})
    &=1-\sum_{i=0}^{M-1}\int_{\mathcal{X}}\pi_ip_i(x)\int_{R_{i}}p_{\hat{x}|x}\dd{\hat{x}} \dd{x}\\
    &=1-\sum_{i=0}^{M-1}\sum_{j=0}^{M-1}\int_{R_j^{*}}\pi_ip_i(x)\int_{R_{i}}p_{\hat{x}|x}\dd{\hat{x}} \dd{x}\label{eqn:split_pe}\\
    &= 1-\sum_{j=0}^{M-1}\int_{R_j^{*}}\sum_{i=0}^{M-1}\pi_ip_i(x)\int_{R_{i}}p_{\hat{x}|x}\dd{\hat{x}} \dd{x}\\
    &=  1-\sum_{j=0}^{M-1}\int_{R_j^{*}}\left(\pi_jp_j(x)+\sum_{i=1,i\neq j}l_{j,i}(x)\int_{R_{i}}p_{\hat{x}|x}\dd{\hat{x}}\right) \dd{x},
\end{align}
where $l_{j,i}(x)=\pi_ip_i(x)-\pi_jp_j(x)$. Now, note that in the region $R_j^*$ we have $l_{j,i}(x)<0$. Therefore, to minimize the error probability, our mapping must satisfy \eqref{eqn:opt_cond1}. Now, substituting these relations back on to  \eqref{eqn:split_pe}, we obtain 
\begin{align}
    p_e(\hat{X})=1-\sum_{j=0}^{M-1}\int_{R_j^{*}}\pi_jp_j(x) \dd{x}=p_e^*.
\end{align}

\section{Proofs of Theorem \ref{thrm:lvl2_err_bound} and Corollary \ref{cor:lvl2_err_bound}}\label{appen:thrm_lvl2_err_bound}
First, note that we can represent $r_e$ as
\begin{align}
    r_e=\sum_{i=0}^{M-1}\int_{\bar{R}_i}\gamma_ir_i(x),\dd{x}
\end{align}
Further, note that
\begin{align}
    P_{\hat{X}_j}(\hat{x})&=\frac{1}{P_j}\int_{R_j^*}p_{\hat{x}|x}\sum_{i=0}^{M-1}\pi_ip_i(x)\dd{x}.
\end{align}
Now, error of the statistical test satisfies
\begin{align}
    p_e(\hat{X})
    =&\sum_{i=0}^{M-1}\int_{\mathcal{X}}\pi_ip_i(x)\int_{\bar{R}_i}p_{\hat{x}|x}\dd{\hat{x}}\dd{x}\\
    =&\sum_{i=0}^{M-1}\int_{\bar{R}_i^*}\pi_ip_i(x)\int_{\bar{R}_i}p_{\hat{x}|x}\dd{\hat{x}}\dd{x}+\sum_{i=0}^{M-1}\int_{R_i^*}\pi_ip_i(x)\int_{\bar{R}_i}p_{\hat{x}|x}\dd{\hat{x}}\dd{x}\\
    \leq&\sum_{i=0}^{M-1}\int_{\bar{R}_i^*}\pi_ip_i(x)\dd{x}+\sum_{i=0}^{M-1}\int_{\bar{R}_i}\int_{R_i^*}\pi_ip_i(x)p_{\hat{x}|x}\dd{\hat{x}}\dd{x}\\
    \leq& p_e^*+\sum_{i=0}^{M-1}P_i\int_{\bar{R}_i}P_{\hat{X}_i}(\hat{x})\dd{\hat{x}}\\
    =&p_e^*+
    \sum_{i=0}^{M-1}P_i\int_{\bar{R}_i}r_i(\hat{x})\dd{\hat{x}}+\sum_{i=0}^{M-1}P_i\left(\int_{\bar{R}_i}P_{\hat{X}_i}(\hat{x})\dd{\hat{x}}-\int_{\bar{R}_i}r_i(\hat{x})\dd{\hat{x}}\right)\\
    \leq&p_e^*+C\sum_{i=0}^{M-1}\gamma_i\int_{\bar{R}_i}r_i(\hat{x})\dd{\hat{x}}+\sum_{i=0}^{M-1}P_i\|P_{\hat{X}_i}-r_i\|_{TV}\\
    \leq&p_e^*+Cr_e+\sum_{i=0}^{M-1}P_i\sqrt{\frac{1}{2}D(P_{\hat{X}_i}||r_i)}\label{eqn:pinsker}\\
    \leq&p_e^*+Cr_e+\sum_{i=0}^{M-1}\sqrt{P_i}\sqrt{\frac{P_i}{2}D(P_{\hat{X}_i}||r_i)}\\
    \leq&p_e^*+Cr_e+\frac{1}{\sqrt{2}}\left(\sum_{i=0}^{M-1}P_iD(P_{\hat{X}_i}||r_i)\right)^{\frac{1}{2}},\label{eqn:cauchy_schwarz}
\end{align}
where \eqref{eqn:pinsker} follows from Pinsker's inequality and \eqref{eqn:cauchy_schwarz} follows from Cauchy-Schwarz inequality.

\section{RDE Function for a Bernoulli Source}\label{appen:Bernouli_1}
We solve the RDE function
\begin{align}
    R(D,\alpha) = \min_{p_{\hat X|X}} & \quad I(X; \hat X) \nonumber\\
    \text{s.t.} & \quad \mathbb{E}[d(X,\hat X)]\leq D \nonumber\\
    & \quad p_e(\hat X)\leq\alpha,
\end{align}
for a binary source. We assume
\begin{align}
    H \in\{0,1\}, \quad P(H=0)=\pi_0, \quad P(H=1)=\pi_1,
\end{align}
where $\pi_0+\pi_1=1$. Conditioned on the hypothesis, let 
\begin{align}
   X|H=0\sim\operatorname{Bern}(a_0),\quad
    X|H=1\sim\operatorname{Bern}(a_1)
\end{align}
Therefore, the marginal distribution of $X$ is also Bernoulli, $X\sim \operatorname{Bern}(a)$ where $a=\pi_0a_0+\pi_1a_1$. We assume that $\hat{X} \in \{0,1\}$, and that the fixed classifier at the receiver decides $\hat H=0$ if $\hat X=0$ and $\hat H=1$ if $\hat X=1$. For a binary reconstruction alphabet, 
\begin{align}
    u\triangleq P(\hat X=1|X=0), \quad v\triangleq P(\hat X=1|X=1),
\end{align}
where $0\leq u,v\leq1$. The probability that the reconstruction equals one is $b\triangleq P(\hat X=1) = (1-a)u+av$. The mutual information is
\begin{align}
    I(X;\hat X) = H_b(b) -(1-a)H_b(u) -a H_b(v).
\end{align}
where $H_b(\cdot)$ is the binary entropy function. Thus, the rate can be written directly as a function of $u$ and $v$. We use Hamming distortion, $d(x,\hat x)=\mathbf{1}\{x\neq\hat x\}$. The average distortion is therefore
\begin{align}
    d(u,v) =& E[d(X,\hat X)] = P(X\neq\hat X) \\
    =&P(X=0)P(\hat X=1|X=0) +P(X=1)P(\hat X=0|X=1) \\
    =&(1-a)u+a(1-v).
\end{align}
Hence, the distortion constraint becomes $(1-a)u+a(1-v)\leq D.$

An error occurs under $H=0$ when $\hat X=1$, and under $H=1$ when $\hat X=0$. Hence, $ p_e = \pi_0P(\hat X=1|H=0) + \pi_1P(\hat X=0|H=1). $ For $H=0$, $ P(\hat X=1|H=0) = (1-a_0)u+a_0v. $ Similarly, $ P(\hat X=0|H=1) = (1-a_1)(1-u)+a_1(1-v). $ Therefore,
\begin{align}
    p_e(u,v) =& \pi_0[(1-a_0)u+a_0v] +\pi_1[(1-a_1)(1-u)+a_1(1-v)].
\end{align}
We write this as $p_e(u,v)=\pi_1+c_0u+c_1v,$ where $c_0=\pi_0(1-a_0)-\pi_1(1-a_1)$ and $c_1=\pi_0a_0-\pi_1a_1$. The probability of error constraint is therefore simply $\pi_1+c_0u+c_1v\leq\alpha.$ Notice that both the distortion constraint and the probability of error constraint are linear in $u$ and $v$. Substituting the previous expressions, the original optimization problem becomes
\begin{align}
    R(D,\alpha)  = \min_{0\leq u,v\leq1} & \quad H_b\big((1-a)u+av\big) -(1-a)H_b(u)-aH_b(v) \nonumber\\
    \text{s.t.} & \quad (1-a)u+a(1-v)\leq D \nonumber\\
    & \quad \pi_1+c_0u+c_1v\leq\alpha.
\end{align}
The mutual information is convex in the conditional distribution $p_{\hat X|X}$ when the source distribution is fixed. Also, both constraints are linear. Therefore, this is a convex optimization problem. Let $\lambda\geq0$ be the multiplier associated with the distortion constraint and let $\beta\geq0$ be the multiplier associated with the probability of error constraint. The Lagrangian $L(u,v,\lambda,\beta)$ is,
\begin{align}
    L(u,v,\lambda,\beta)=&H_b\big((1-a)u+av\big) -(1-a)H_b(u)-aH_b(v) \nonumber\\
    &+\lambda\left[(1-a)u+a(1-v)-D\right] +\beta\left[\pi_1+c_0u+c_1v-\alpha\right].
\end{align}
For the derivative calculation, we use natural logarithms, which only changes the scaling of $\lambda$ and $\beta$, and does not change the optimal $u$ and $v$. Define $\operatorname{logit}(x)=\log\frac{x}{1-x}$ and $\sigma(x)=\frac{1}{1+e^{-x}}.$ Let $b=(1-a)u+av$. For $0<u,v<1$, differentiating the Lagrangian with respect to $u$ gives $ \operatorname{logit}(u) = \operatorname{logit}(b) -\lambda-\beta\frac{c_0}{1-a}$; and differentiation with respect to $v$ gives $ \operatorname{logit}(v) = \operatorname{logit}(b)+\lambda-\beta\frac{c_1}{a}.$ Thus,
\begin{align}
    u&= \sigma\left(\operatorname{logit}(b) -\lambda -\beta\frac{c_0}{1-a}\right), \\
    v&=
    \sigma\left(\operatorname{logit}(b)+\lambda-\beta\frac{c_1}{a}\right).
    \label{eqn:bern_main_uv}
\end{align} 
 The multipliers also satisfy $\lambda\left[(1-a)u+a(1-v)-D\right]=0$ and $\beta\left[\pi_1+c_0u+c_1v-\alpha\right]=0.$

\subsection*{Zero Rate Region}
A possibility is $R(D,\alpha)=0$. The mutual information vanishes if and only if $X$ and $\hat X$ are independent. For the binary channel considered here, this means $u=v=q$, $0\leq q\leq 1$. For such a channel, $b=q$ and $I(X;\hat X)=0$. The distortion becomes $d(q)=(1-a)q+a(1-q)=a+(1-2a)q$. The classification error becomes $p_{e}(q)=\pi_1+(c_0+c_1)q$. Since $c_0+c_1=\pi_0-\pi_1$ we obtain $p_e(q)=    \pi_1+(\pi_0-\pi_1)q = \pi_0q+\pi_1(1-q)$. Therefore, $R(D,\alpha)=0$ if and only if there exists a $q\in[0,1]$ such that $a+(1-2a)q\leq D$ and $\pi_1+(\pi_0-\pi_1)q\leq \alpha$. 

In the absence of a distortion constraint, a zero-rate classifier can always choose a constant reconstruction corresponding to the more probable hypothesis. Hence, a zero-rate reconstruction satisfying only the error constraint exists whenever $\alpha\geq\min\{\pi_0,\pi_1\}$.

\subsection*{Case 1: Only Distortion Constraint is Active}
Suppose $\lambda>0$, $\beta=0$. The problem then reduces to the Bernoulli RD problem. For $0\leq D<\min\{a, 1-a\}$ the optimal marginal is $b=\tfrac{a-D}{1-2D}$. The optimal $u_D$ and $v_D$ are 
\begin{align}
    u_D=\frac{D(a-D)}{(1-a)(1-2D)},\quad v_D=\frac{(1-D)(a-D)}{a(1-2D)}.
\end{align}
They satisfy $(1-a)u_D+a(1-v_D)=D$. The corresponding rate is 
\begin{align}
    R_D(D)=H_b(a)-H_b(D), \quad 0\leq D<\min\{a,1-a\}.
\end{align}
This channel is also optimal for the RDE problem
whenever it satisfies the classification-error constraint,
\begin{align}
    \pi_1+c_0u_D+c_1v_D\leq\alpha.
    \label{eqn:rd-error-condition}
\end{align}
Thus, whenever \eqref{eqn:rd-error-condition} holds, $R(D,\alpha)=H_b(a)-H_b(D)$. For $D\geq\min\{a,1-a\}$, the Bernoulli RD function is zero.

\subsection*{Case 2: Only Probability of Error Constraint is Active}
Now, suppose $\lambda=0$, $\beta>0$. Then, \eqref{eqn:bern_main_uv} become
\begin{align}
    \operatorname{logit}(u_E) &= \operatorname{logit}(b) -\beta\frac{c_0}{1-a}, \\
    \operatorname{logit}(v_E)&= \operatorname{logit}(b)-\beta\frac{c_1}{a}.
\end{align}
Define $t_0\triangleq\exp(-\beta\tfrac{c_0}{1-a})$ and $t_1\triangleq\exp(-\beta\tfrac{c_1}{a})$. Then, 
\begin{align}
    u_E=\frac{b t_0}{1-b+b t_0}, \quad v_E=\frac{b t_1}{1-b+b t_1}.
\end{align}
Using the relation $b = (1-a)u_E+av_E$, we have
\begin{align}
    b = \frac{1-(1-a)t_0-at_1}{(1-t_0)(1-t_1)}.
    \label{eqn:rde-b}
\end{align}
This expression applies whenever the denominator is nonzero and is in $(0,1)$. In some cases where \eqref{eqn:rde-b} is not found, $b$ is obtained directly from the fixed-point equation $b=(1-a)u_E+av_E$. The Lagrange multiplier $\beta$ is selected so that the active error constraint satisfies $\pi_1+c_0u_E+c_1v_E=\alpha$. Thus, this case reduces to a one-dimensional equation in $\beta$. The resulting solution is valid provided its distortion satisfies $(1-a)u_E+a(1-v_E)\leq D$. Then, the rate is
\begin{align}
    R(D,\alpha)= H_b(b) -(1-a)H_b(u_E) -aH_b(v_E).
\end{align}
As $\beta$ increases, the optimal probability of error is non-increasing. Hence, the appropriate value of $\beta$ can be determined numerically using a one-dimensional monotone search.

\subsection*{Case 3: Both Constraints are Active}
Suppose $\lambda>0$, $\beta>0$. Then the complementary slackness condition gives $(1-a)u+a(1-v)=D$ and $\pi_1+c_0u+c_1v=\alpha$. Equivalently $(1-a)u-av=D-a$ and $c_0u+c_1v=\alpha-\pi_1$. Then, let $\Delta\triangleq(1-a)c_1+ac_0$, if $\Delta\neq0$, \eqref{eqn:bern_main_uv} has a unique solution
\begin{align}
    u^\star=\frac{c_1(D-a)+a(\alpha-\pi_1)}{\Delta}, \quad
    v^\star&=\frac{(1-a)(\alpha-\pi_1)-c_0(D-a)}{\Delta}.
    \label{eqn:uv-both}
\end{align}
The corresponding reconstruction probability is $b^\star = (1-a)u^\star+av^\star$, and therefore
\begin{align}
    R(D,\alpha) = H_b(b^\star) -(1-a)H_b(u^\star) -aH_b(v^\star)
\end{align}
Solution in \eqref{eqn:uv-both} is admissible only if $0\leq u^\star\leq1$ and $0\leq v^\star\leq1$. For an interior solution, the multipliers must also satisfy $\lambda>0$ and $\beta>0$. Define $A\triangleq\operatorname{logit}(b^\star) - \operatorname{logit}(u^\star)$ and $B\triangleq\operatorname{logit}(v^\star) - \operatorname{logit}(b^\star)$. Then, $A=\lambda + \beta\tfrac{c_0}{1-a}$ and $B=\lambda - \beta\tfrac{c_1}{a}$. Thus,
\begin{align}
    \lambda = \operatorname{logit}(b^\star)-\operatorname{logit}(u^\star) - \beta\frac{c_0}{1-a}, \ 
    \beta = \frac{a(1-a)}{\Delta}\big[2\operatorname{logit}(b^\star)-\operatorname{logit}(u^\star)-\operatorname{logit}(v^\star)\big].
\end{align}

Therefore, the optimum solution is at the intersection of the  constraint lines defined by $(1-a)u-av=D-a$ and $c_0u+c_1v=\alpha-\pi_1$, as long as that intersecting point is inside the allowed $0$ to $1$ range and the multipliers are nonnegative. If $\Delta=0$, i.e., $\Delta = (1-a)c_1+ac_0=0$, the above constraint lines are parallel. If the equations $(1-a)u-av=D-a$ and $c_0u+c_1v=\alpha-\pi_1$ are inconsistent, the two constraints cannot be active simultaneously. If the two equations are proportional and consistent, the problem then reduces to minimizing the convex mutual information function along a single line segment. For example using the $(1-a)u-av=D-a$, we get $v=\tfrac{(1-a)u+a-D}{a}$. Then, we minimize
\begin{align}
    I_D(u)\triangleq H_b\Big((1-a)u+a\tfrac{(1-a)u+a-D}{a}\Big) - (1-a)H_b(u)-aH_b\Big(\tfrac{(1-a)u+a-D}{a}\Big)
\end{align}
over the interval $0\leq u\leq1$ and $0\leq\tfrac{(1-a)u+a-D}{a}\leq1$. Since this is one dimensional convex optimization problem its minimum occurs either at unique stationary point or at an endpoint.

\subsection*{Final Characterization}

For arbitrary $\pi_0,\pi_1\geq0$ where $\pi_0+\pi_1=1$ and $1\geq a_0,a_1\geq 0$ where $a=\pi_0a_0+\pi_1a_1$. The RDE function can be determined as follows.
If $D$ and $\alpha$ are not in the feasible set, the $R(D,\alpha)$ function is not defined. If there exists $q\in[0,1]$ such that $a+(1-2a)q\leq D$ and $\pi_1+(\pi_0-\pi_1)q\leq \alpha$ then $R(D,\alpha)=0$. Otherwise $R(D,\alpha)>0$ and the optimizer is obtained via
\begin{align}
    (u^*,v^*)=
    \begin{cases}
        (u_D,v_D), & \text{if }\lambda>0,\beta=0,\\
        (u_E,v_E), & \text{if } \lambda=0,\beta>0,\\
        \big(\tfrac{c_1(D-a)+a(\alpha-\pi_1)}{\Delta}, \tfrac{(1-a)(\alpha-\pi_1)-c_0(D-a)}{\Delta}\big), & \text{otherwise.}
    \end{cases}
\end{align}
In all cases, once the optimal pair $(u^*,v^*)$ is found, the rate is
\begin{align}
    R(D,\alpha)=H_b((1-a)u^*+av^*)-(1-a)H_b(u^*)-aH_b(v^*).
\end{align}
Thus, for arbitrary $\pi_0,\pi_1,a_0,a_1$, the RDE function is characterized by a finite set of convex optimization branches. The distortion only and both active branches admit closed-form solutions, while the error-only branch reduces to a one-dimensional equation for the multiplier $\beta$.

\begin{figure}
    \centering
    \includegraphics[trim=0 0 0 0cm, clip,width=0.5\linewidth]{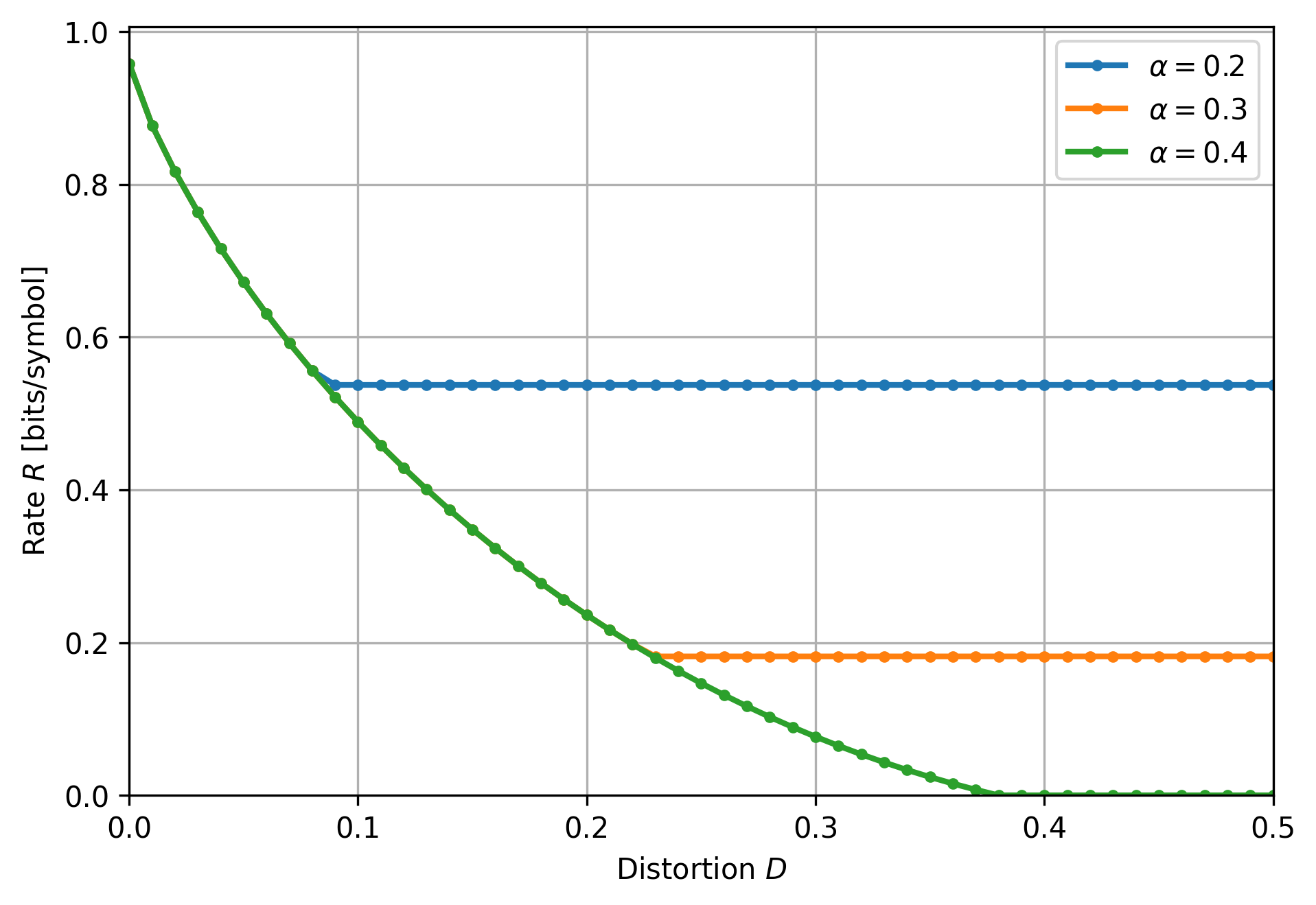}
    \caption{Rate–distortion curves for different classification error constraints $\alpha \in \{0.2,0.3,0.4\}$ with $X|H_0\sim \text{Bern}(0.1)$, $X|H_1\sim \text{Bern}(0.8)$, $\pi_0=0.6$ and $\pi_1=0.4$.}
    \label{fig:rdc} 
\end{figure}

\section{Experimental Setup for Section \ref{sec:num_res}}\label{appen:xtra_discrete}
In this section, we present the distributions $p_i(x)$ and $q_i(x)$ used for the binary hypothesis experiments discussed in Section \ref{sec:num_res}. We consider a quaternary alphabet on $\{0,1,2,3\}$ and consider $p_0(x)$ to be a skewed distribution concentrated on $\{0,1\}$ whereas $p_1(x)$ is a skewed distribution concentrated on $\{2,3\}$. MAP rule is applied on them to find the optimal decision regions $R_i^*$. These optimal decision regions yield that $p_e^*=0.12$. We consider that the statistical test used at the receiver is also MAP. However, it was designed for using distributions $q_i(x)$ defined in Table \ref{tab:quat_dist}. These decisions yield $r_e=0.03$. As seen in Table \ref{tab:quat_dist}, there is a clear mismatch between the optimal decision regions at the transmitter with that used at the receiver.
\begin{table}[!h]
    \centering
    \begin{tabular}{|c|c|c|c|c|c|c|}
    \hline
         $\mathcal{X}$&$p_0(x)$&$p_1(x)$&$q_0(x)$&$q_1(x)$&Tx DR&Rx DR  \\
    \hline
        0&0.68 &0.02 &0.01 &0.3&$R_0^*$&$R_1$\\
    \hline
        1&0.2 &0.1 &0.95 &0.01&$R_0^*$&$R_0$\\
    \hline
        2&0.1 &0.2 &0.03 &0.4&$R_1^*$&$R_1$\\
    \hline
        3&0.02 &0.68 &0.01 &0.29&$R_1^*$&$R_1$\\
    \hline
    \end{tabular}
    \caption{Distributions and decision regions (DR).}
    \label{tab:quat_dist}
\end{table}

\section{Architecture and Training Parameters for the Experiments in Section~\ref{sec:img_exp}}
\label{appen:img_exp_params}

\subsection{Datasets and Classifiers}
Both domains are rendered as $32\times32$ RGB images: SVHN \citep{svhn} in its native resolution and MNIST \citep{mnist} replicated across the three channels and resized, so that $X$ and $Y$ share the support $\mathcal{X}$ required by the problem formulation. We use $8000$ training, $1500$ validation and $3000$ test images per domain, resampled so that the realized class frequencies match the uniform priors $\pi_i=\sigma_i=1/M$.

The receiver's test $g_Y$ and the transmitter's estimator $g_X$ are both ResNet-9 networks with base width $64$ ($6.57$M parameters), trained with AdamW at a learning rate of $2\times10^{-3}$, weight decay $5\times10^{-4}$, batch size $512$, cosine schedule with warmup, label smoothing $0.05$, dropout $0.05$, random-crop and flip augmentation, for at most $100$ epochs with early stopping on validation accuracy (patience $8$). Each is trained on its own domain only: $g_Y$ on MNIST under $\sigma_i$, $g_X$ on SVHN under $\pi_i$. Their accuracies are reported in Table \ref{tab:img_classifiers}. Neither network is ever placed in the computational graph of the encoder or decoder, $g_Y$ is used only to evaluate $p_e(\hat{X})$ after training, and $g_X$ is used only to produce $J$ and the feature space of the distortion measure described below.

\begin{table}[h]
    \centering
    \begin{tabular}{|l|c|c|c|}
    \hline
        & train & validation & test \\
    \hline
        $g_X$ (SVHN) & 1.000 & 0.938 & 0.921 \\
    \hline
        $g_Y$ (MNIST) & 1.000 & 0.993 & 0.992    \\
    \hline
    \end{tabular}
    \caption{Accuracy of the transmitter's estimator and the receiver's test, each on its own domain.}
    \label{tab:img_classifiers}
\end{table}

\subsection{Encoder, Decoder and Rate}
The encoder and decoder are convolutional and are given in Table \ref{tab:architecture}. The encoder maps a $32\times32\times3$ image to a $128$-dimensional latent vector, which is split into $L$ tokens of dimension $128/L$ and quantized against a learned codebook of $K$ entries by nearest-neighbor assignment, with straight-through gradients and a commitment penalty of weight $0.25$. The rate is therefore fixed by the geometry of the quantizer as $R=L\log_2K$ bits per image and does not have to be estimated or penalized. When $J$ is transmitted, it is charged $H(J)=\log_2 M\approx3.32$ bits, which is added to the rates reported in Section \ref{sec:img_exp}. The four rate points used are listed in Table \ref{tab:img_rates}. The transmitted $J$ enters the network as a learned $32$-dimensional embedding which modulates both the encoder and the decoder through FiLM layers whose output projection is initialized to zero, so that conditioning begins as the identity map. Following \citep{blau_rdp}, uniform noise of half the code spacing is added to the quantized representation during training so that the decoder is stochastic, which is necessary for good perceptual quality at low rates.

\begin{table}[h]
    \centering
    \begin{tabular}{c|c}
        Encoder & Decoder \\
        \hline
        $32\times32\times3$ & $L\log_2K$ bits \\
        Conv $4\times4$, stride 2: $16\times16\times32$ & FC: $128\to256$ \\
        Conv $4\times4$, stride 2: $8\times8\times64$ & FiLM($J$) \\
        AvgPool: $4\times4\times64$ & FC: $256\to 8\times8\times64$ \\
        FC: $1024\to256$ & ConvT $4\times4$, stride 2: $16\times16\times32$ \\
        FiLM($J$) & ConvT $4\times4$, stride 2: $32\times32\times32$ \\
        FC: $256\to128$ & Conv $3\times3$: $32\times32\times3$, sigmoid \\
        VQ: $L$ tokens, $K$ codewords & \\
    \end{tabular}
    \caption{Encoder and decoder architecture.}
    \label{tab:architecture}
\end{table}

\begin{table}[h]
    \centering
    \begin{tabular}{|c|c|c|c|c|}
    \hline
        $L$ & $K$ & token dim. & $R$ [bits] & $R+H(J)$ [bits] \\
    \hline
        8 & 8 & 16 & 24 & 27.32 \\
    \hline
        8 & 16 & 16 & 32 & 35.32 \\
    \hline
        16 & 32 & 8 & 80 & 83.32 \\
    \hline
        32 & 64 & 4 & 192 & 195.32 \\
    \hline
    \end{tabular}
    \caption{Rate points. The rate is exact by construction, $R=L\log_2K$.}
    \label{tab:img_rates}
\end{table}

\subsection{Distortion Measure}
We use the combination of squared error and a deep feature distance of \cite[Eqn.~(11)]{blau_rdp}:
\begin{align}
    \Delta(x,\hat{x})=\|x-\hat x\|^2+\alpha\|\Psi(x)-\Psi(\hat x)\|^2,
    \label{eqn:blau_dist}
\end{align}
with $\alpha=1$. As in \citep{blau_rdp}, the feature map $\Psi$ is an intermediate layer of a trained classifier rather than a generic network, because the term is only meaningful if the feature space carries semantic information. We take $\Psi$ to be the next to last representation of the transmitter's estimator $g_X$. We have two main reasons for doing this. First, $g_X$ is trained on the transmitter domain data that the transmitter owns, therefore, using it discloses nothing about the receiver's test and keeps the scheme within Level 2 and 3. Second, the depth of the layer matters more than the accuracy of the classifier since we observed that measuring the ratio of the mean squared feature distance between images of different classes to that between images of the same class. A feature space is useful in \eqref{eqn:blau_dist} only if the distance reflects class differences to choose the layer we measured, for each candidate $\Psi$,
\begin{align}
    \mathrm{sep}(\Psi)=\frac{\mathbb{E}\big[\|\Psi(X)-\Psi(X')\|^2\big]}
                            {\mathbb{E}\big[\|\Psi(X)-\Psi(X'')\|^2\big]},
    \label{eqn:sep_metric}
\end{align}
where $X$ and $X''$ are drawn from the same class and $X$ and $X'$ from different classes. The ratio is not interpretable on its own, since raw pixel distance already separates classes to some extent. The identity function $\Psi(x)=x$ is therefore measured on the same images as baseline, and what matters is the ratio relative to it. Table \ref{tab:feature_layers} shows the results for $g_X$ at $0.921$ test accuracy. An early layer scores $1.003$ against $0.099$ for raw pixels, adding almost nothing over pixel distance, whereas the last layer representation scores $1.543$. We therefore use the deepest convolutional representation.

\begin{table}[h]
    \centering
    \begin{tabular}{|l|c|c|c|}
    \hline
        feature map $\Psi$ & dimension & $\mathrm{sep}(\Psi)$ & relative to pixels \\
    \hline
        raw pixels (identity)        & 3072 & 0.990 & 1.000 \\
    \hline
        $g_X$, early                 & 65536 & 1.003 & 1.013 \\
    \hline
        $g_X$, middle                & 32768 & 1.041 & 1.051 \\
    \hline
        $g_X$, penultimate (used)    & 8192 & 1.528 & 1.543 \\
    \hline
    \end{tabular}
    \caption{Class separation \eqref{eqn:sep_metric} of each candidate feature space.}
    \label{tab:feature_layers}
\end{table}

\subsection{Perception Measure}
The divergence $\delta(\cdot,\cdot)$ in \eqref{eqn:img_lagrangian} is the Wassertein-1 distance, estimated as in \citep{blau_rdp} by a critic trained with gradient penalty \citep{wgan_gp}
\begin{align}
    \delta(P_{\hat{X}_i},P_{Y_i})=\max_{h\in\mathcal{F}}\ \e[h(Y_i)]-\e[h(\hat{X}_i)],
\end{align}
where $\mathcal{F}$ is realized by the network in Table \ref{tab:critic}. A single critic serves all $M$ branches, the class embedding is projected onto the image features and added to the scalar output, in the manner of a projection discriminator, and the critic is trained on the pooled batch with the branch labels. We use $5$ critic updates per generator update, a gradient penalty weight of $10$, and Adam with learning rate $10^{-4}$ and $(\beta_1,\beta_2)=(0.5,0.9)$. The critic is trained at every step, including at $\lambda=0$ where it does not influence the objective, and is refit for a further $300$ steps with the encoder and decoder frozen before the perception index is reported, so that the values compared across $\lambda$ are measured by critics at comparable convergence.

\begin{table}[h]
    \centering
    \begin{tabular}{c|c}
        Layer & Output \\
        \hline
        Input & $32\times32\times3$ \\
        Conv $4\times4$, stride 2, LeakyReLU(0.2) & $16\times16\times64$ \\
        Conv $4\times4$, stride 2, LeakyReLU(0.2) & $8\times8\times128$ \\
        AvgPool, Flatten & $2048$ \\
        FC, LeakyReLU(0.2) & $256$ \\
        FC $+$ class projection & $1$ \\
    \end{tabular}
    \caption{Critic used to estimate the perception index.}
    \label{tab:critic}
\end{table}

\begin{figure}[t]
    \centering
    \includegraphics[width=1\linewidth]{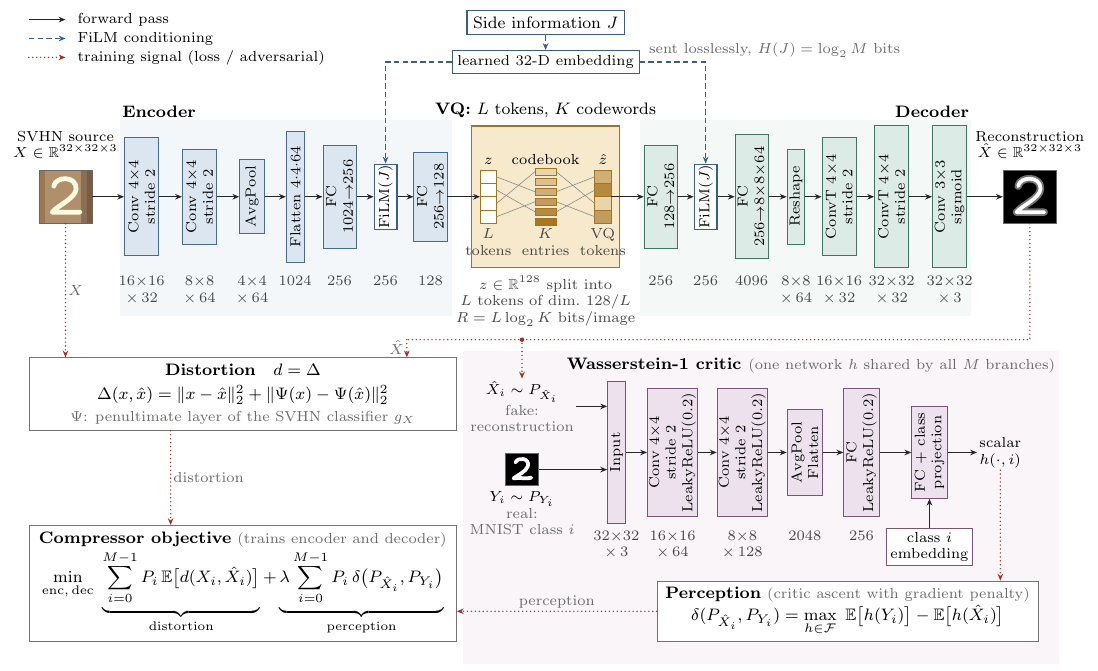}
    \caption{System architecture diagram.}
    \label{fig:architecture}
\end{figure}

\subsection{Training of the Compression Scheme}
Each operating point is a separate encoder-decoder pair trained for $300$ epochs with Adam at a learning rate of $10^{-3}$ and a batch size of $800$, on the $8000$ transmitter domain training images. The reference batches for the perception term are drawn from the receiver domain training split for the $P_{Y_i}$ configuration and from the transmitter domain split for the $P_{X_i}$ control; the test split is never used during training.

Following the protocol of \citep{blau_rdp}, for each rate, we first train the distortion-only model ($\lambda=0$) and every $\lambda>0$ model is initialized from it and trained for the same number of further epochs. Asking an untrained codec to match a distribution while it is still learning to reconstruct rarely converges within a practical budget, and this warm start also removes the initialization as a source of variation between the points of a sweep.

Finally, we note that the numerical values of $\lambda$ are not comparable across divergences unless the two terms of \eqref{eqn:img_lagrangian} are on a common scale. The distortion above is a per-pixel mean of order $10^{-2}$, whereas the critic estimate is of order $10$, thus a literal $\lambda$ would suppress the distortion term entirely. We therefore measure the magnitude of both terms over a short warmup and apply $\lambda$ to their ratio, so that $\lambda=1$ weights the two terms equally, the $\lambda$ grid used is $\{0, 0.1,0.5,1,1.5,2,2.5,3,3.5,4,4.5,5,10,20,35,50,75,90,100,300\}$. Reported distortion and perception values are the achieved quantities and are unaffected by this convention.

\section{Additional Results}\label{appen:xtra_results}
\subsection{RDP Surface}
\begin{figure}[!h]
    \centering
    \begin{subfigure}[t]{0.45\linewidth}
            \vspace{0pt}
            \centering
            \includegraphics[width=1.23\linewidth]{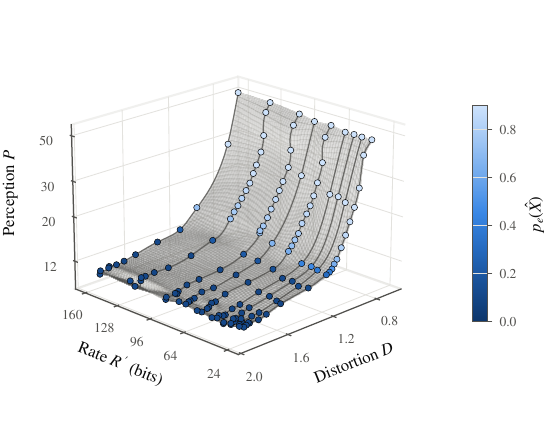}
            \caption{rate}
            \label{fig:3d_rdp_surface}
        \end{subfigure}
        ~
        \begin{subfigure}[t]{0.45\linewidth}
            \vspace{0pt}
            \centering
            \includegraphics[trim={0cm 0cm 0cm 0cm}, clip,width=\linewidth]{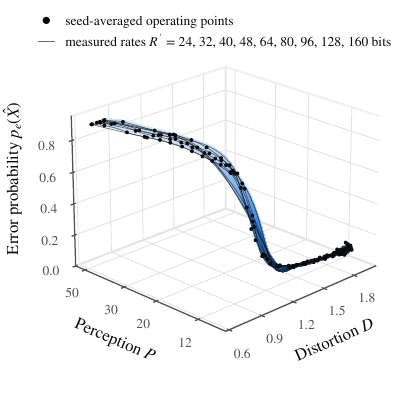}
            \caption{$p_e(\hat{X})$}
            \label{fig:3d_edp_surface}
        \end{subfigure}
        \caption{Variation of rate and error with distortion and perception (Level 2 Tx $J$).}
\end{figure}

Fig.~\ref{fig:3d_rdp_surface} illustrates the tradeoff between rate, distortion and perception for Level 2 with lossless transmission of $J$. This surface was obtained by sweeping the Lagrangian described in Section \ref{sec:img_exp}. As illustrated, the classification error decreases towards the low perception and high distortion regime similar to our results with discrete alphabets. This trend is further highlighted in the error, distortion and perception curve depicted in Fig.~\ref{fig:3d_edp_surface}.

\subsection{Reconstruction Results}
\begin{figure}[htp]
    \centering
    \begin{subfigure}[b]{\textwidth}
        \centering
        \includegraphics[trim=0cm 0cm 0cm 0.5cm, clip,width=.9\textwidth]{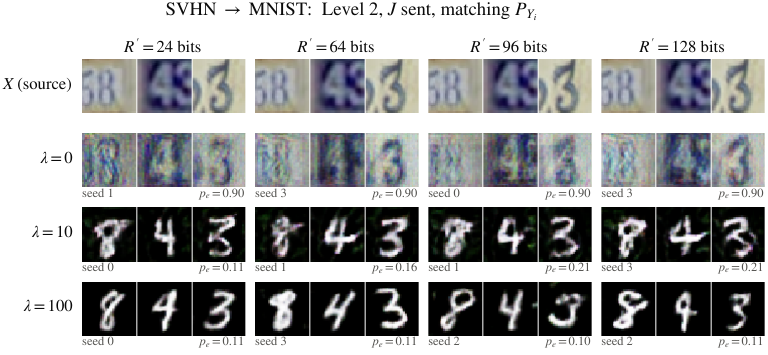}
        \caption{Level 2 Tx $J$.}
        \label{fig:recons_L2_sendJ_Py}
    \end{subfigure}
    
    \bigskip
    
    \begin{subfigure}[b]{\textwidth}
        \centering
        \includegraphics[trim=0cm 0cm 0cm 0.5cm, clip,width=.9\textwidth]{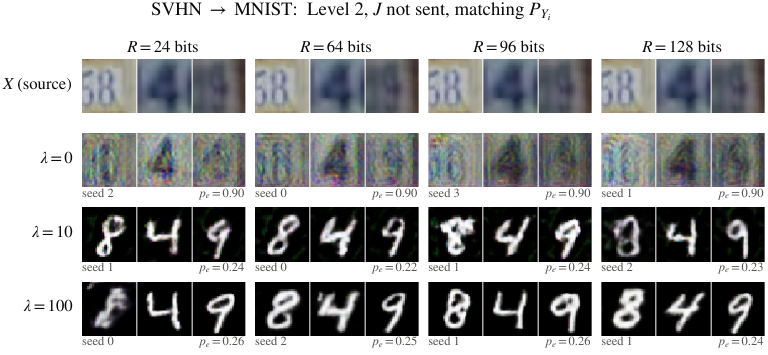}
        \caption{Level 2 No $J$.}
        \label{fig:recons_L2_noJ_Py}
    \end{subfigure}
    \caption{Reconstruction results for Level 2.}
    \label{fig:recons_L2_Py}
\end{figure}

\begin{figure}[p!]
\begin{minipage}[t]{\textwidth}
        \centering
        \includegraphics[trim=0cm 0cm 0cm 0.5cm, clip,width=.9\textwidth]{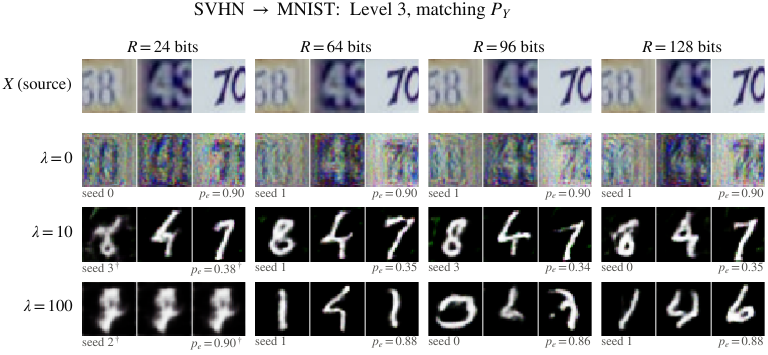}
        \caption{Reconstructions for Level 3.}
        \label{fig:recons_L3_Py}
    \end{minipage}
    
\bigskip

\begin{minipage}[t]{\textwidth}
        \centering
        \includegraphics[trim=0cm 0cm 0cm 0.5cm, clip,width=.9\textwidth]{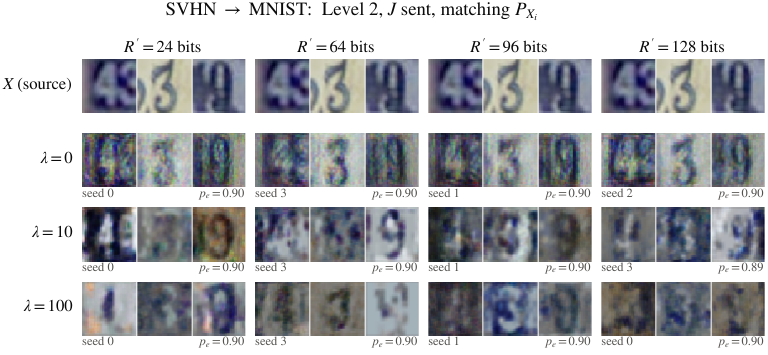}
        \caption{Level 2, Tx J, match $P_X$.}
        \label{fig:recons_L2_sendJ_Px}
    \end{minipage}
\end{figure}

Fig.~\ref{fig:recons_L2_Py} and Fig.~\ref{fig:recons_L3_Py} depict sample reconstructions obtained at Level 2 and Level 3, respectively. We see that by matching on to the MNIST data distribution, the decoder generated images resembling samples from the MNIST dataset that is close to our source sample from SVHN dataset. We see that in Level 2, the reconstructions that are obtained at higher $\lambda$ values (i.e., tighter perception constraint) resembles samples that are closer to MNIST dataset and hence the decrease in error. However, for Level 3, we observe that the reconstructions at moderate $\lambda$ values generate the best reconstructions from the MNIST dataset. This is because, as $\lambda$ increases, our distortion constraint also increases, thus allowing the decoder to generate reconstructions independently of the actual source sample. Even though the basic features of the MNIST dataset is present in these reconstructions, they no longer contain any useful information from the source sample. Hence, the decrease in the accuracy in Level 3 at high $\lambda$ values. 

We also test how the classification accuracies would vary when we match on to the source distribution following the Level 2 configuration where we transmit $J$ losslessly. Here, the divergence term in  \eqref{eqn:img_lagrangian} was replaced with $\sum_{i=0}^{M-1}P_i\,\delta(P_{\hat{X}_i},P_{X_i})$ and the same training process described in Section \ref{sec:img_exp} was carried out. Fig.~\ref{fig:recons_L2_sendJ_Px} illustrates the reconstructions obtained using the above process. As seen, the reconstructions do resemble samples from the SVHN dataset. However, since the classifier was trained on the MNIST dataset, the error constantly remains around $0.90$ regardless of the $\lambda$ value that is used.

\end{document}